\documentclass[final,3p,times,twocolumn]{elsarticle}

\pdfoutput=1

\usepackage{lineno}

\usepackage{graphicx}
\usepackage{amssymb}
\journal{Nucl. Instr. and Meth. in Phys. Res. A}

\begin{document}

\begin{frontmatter}

%\title{First tests of the applicability of $\gamma$-ray imaging for enhanced sensitivity neutron capture cross section measurements at CERN n\_TOF}
\title{First tests of the applicability of $\gamma$-ray imaging for background discrimination in time-of-flight neutron capture measurements}

\author{D.L.~P\'erez Mag\'an$^{a}$, L.~Caballero-Ontanaya$^{a}$, C.~Domingo-Pardo$^{a,*}$, J.~Agramunt-Ros$^a$, F.~Albiol$^{a}$, A.~Casanovas$^{b}$, A.~Gonz\'alez$^{c}$, C.~Guerrero$^{d}$, J.~Lerendegui-Marco$^{d}$, A.~Tarife\~no-Saldivia$^{a,b}$ and the n\_TOF Collaboration}
\address[1]{IFIC, CSIC-Universidad de Valencia, E-46071 Valencia, Spain}
\address[2]{Institut de T\`ecniques Energ\`etiques -  Departament de F\'{\i}sica i Enginyeria Nuclear, Universitat Polit\`ecnica de Catalunya, 08028, Barcelona, Spain}
\address[3]{Institute for Instrumentation in Molecular Imaging, I3M-CSIC 46022 Valencia, Spain}
\address[4]{Dto. de F\'{\i}sica At\'omica, Molecular y Nuclear, Universidad de Sevilla, 41012 Sevilla, Spain}

\fntext[myfootnote]{Corresponding author domingo@ific.uv.es}

%\maketitle % Insert t	\thispagestyle{fancy} % All pages have headers and footers
%----------------------------------------------------------------------------------------
\begin{abstract}
		
In this work we explore for the first time the applicability of using $\gamma$-ray imaging in neutron capture measurements to identify and suppress spatially localized background. For this aim, a pinhole gamma camera is assembled, tested and characterized in terms of energy and spatial performance. It consists of a monolithic CeBr$_3$ scintillating crystal coupled to a position-sensitive photomultiplier and readout through an integrated circuit AMIC2GR. The pinhole collimator is a massive carven block of lead.
A series of dedicated measurements with calibrated sources and with a neutron beam incident on a $^{197}$Au sample have been carried out at n\_TOF, achieving an enhancement of a factor of two in the signal-to-background ratio when selecting only those events coming from the direction of the sample.
\end{abstract}


\end{frontmatter}

%----------------------------------------------------------------------------------------
%	ARTICLE CONTENTS
%----------------------------------------------------------------------------------------

\section{Introduction}

A very active field of research in nuclear astrophysics is the study of heavy element ($>$Fe) nucleosynthesis in different stellar environments. In environments where the neutron density is low, heavy elements are believed to be formed through chains of slow neutron-captures, known as $s$-process reactions, with iron as a starting point \cite{b2fh}. Apart from the well known half-lives of the unstable isotopes along the chain, the fundamental nuclear physics input for a quantitative understanding of the physical conditions operating in the $s$ process is the neutron capture cross section of the involved nuclei~\cite{kaeppeler11}. Neutron capture time-of-flight measurements can provide the relevant information over the full stellar energy range of interest, from few eV up to several hundreds of keV. Presently, this is done at facilities such as IRMM-Gelina (Belgium), JPARC-ANNRI (Japan), LANL-DANCE (USA) and CERN n\_TOF (Switzerland). 

One of the dominant sources of background in this kind of measurements arises from incident neutrons that, after being scattered by the sample or the vacuum windows along the beam-line, are captured by the surrounding materials or in the walls of the experimental hall.  Such contaminant reactions lead to emission of $\gamma$ rays that are hard to distinguish from those produced in the sample (see e.g. Fig.~6 in~\cite{Zugec14}).

A way of enhancing the signal-to-background (S/B) ratio could be the use of a $\gamma$-ray detector with imaging capability, so called i-TED~\cite{iTED}. Events coming from another direction than the target could be discriminated by exploiting the $\gamma$-ray imaging capability of the detector. In principle, this should filter out localized background events. 
%In addition, since any $\gamma$-ray imaging technique inevitably implies a low $\gamma$-ray detection efficiency, the pulse-height weighting technique needs to be applied in order to avoid the capture efficiency dependency on the particular characteristics of the prompt g-ray cascade path.

To explore this idea the work is organized in two parts. The first one concerns the preparation of a pinhole-based gamma camera. This $\gamma$-ray imager is based on a bulky lead collimator and therefore, its performance for neutron capture measurements is not expected to be as good, or even comparable, to the one of the i-TED Compton modules described in Ref.\cite{iTED}. However, the rather simple device we use still allows us to implement and explore the applicability of gamma imaging under the severe background conditions of this kind of measurement, and to test its basic components under real experimental conditions. This detection system is assembled and tested, as reported in Sec.~\ref{sec:psd}. In Sec.~\ref{sec:psd characterization} its spatial sensitivity is probed by using spatial reconstruction algorithms to demonstrate that we can successfully recover the position of pointlike radioactive sources. The second part, which is described in Sec.~\ref{sec:measurementsatntof}, is carried out at CERN n\_TOF (Geneva). After a brief introduction to the facility (Sec.\ref{ssec:facility}) we describe the experimental set-up used in the present study (Sec.~\ref{ssec:setupatntof}), the detector data-readout scheme (Sec.\ref{ssec:daq}) and the four different measurements performed (Sec.~\ref{sec:measurements}). The data analysis and the interpretation of the results is discussed in Sec.~\ref{sec:analyse}. Finally, the main conclusions and future prospects of this technique are summarized in Sec.~\ref{sec:outlook}.

\section{Position-sensitive detection system}\label{sec:psd}
\begin{figure}[!htbp]
\centering
\includegraphics[width=\columnwidth]{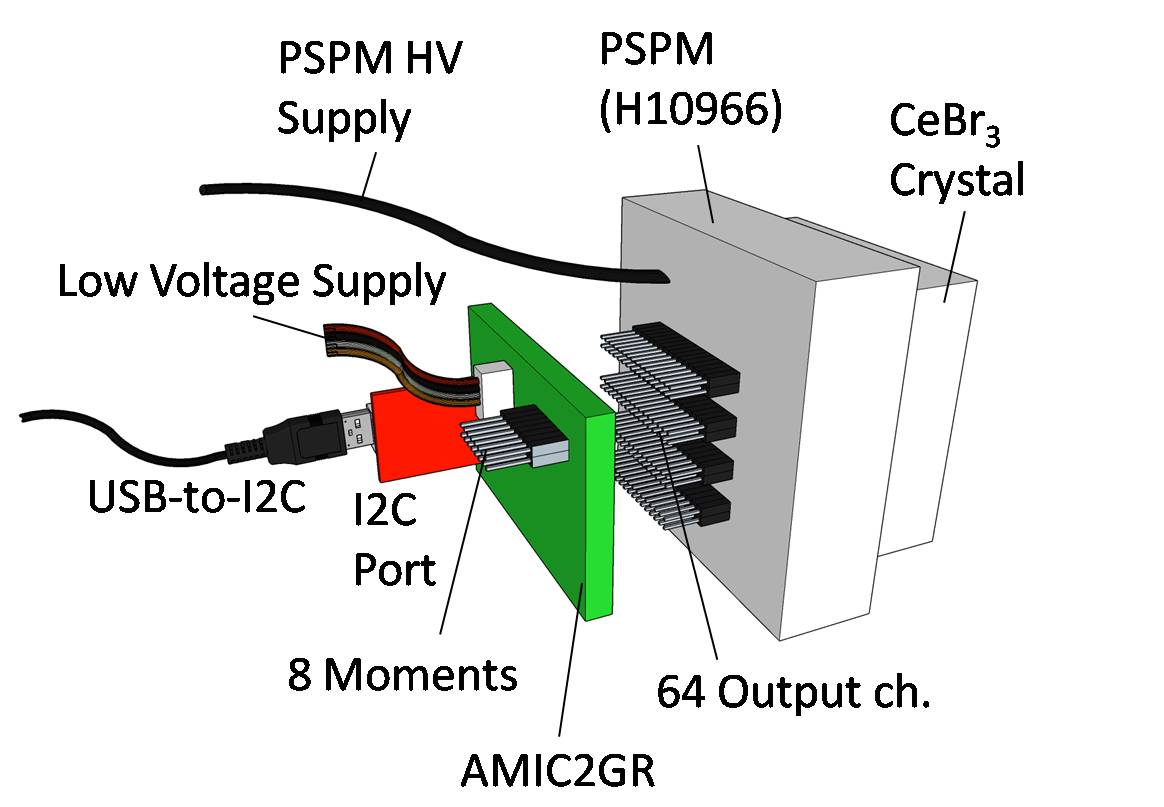} 
\caption{Schematic view of the position-sensitive detection system. The pixelated photomultiplier is optically coupled to a monolithic CeBr$_3$ crystal and readout using an integrated circuit AMIC2GR (green).} \label{fig:detector}
\end{figure}

The $\gamma$-ray detector consists of a monolithic CeBr$_3$ scintillating crystal with a size of 51$\times$51$\times$15 mm$^3$ coupled to a position-sensitive photomultiplier (PSPM) from Hamamatsu Photonics (H10966A100). The crystal choice is driven by availability and, in principle, other materials such as LaCl$_3$ would perform better in terms of neutron sensitivity. The PSPM features 8$\times$8 pixels over an effective photocathode area of 48$\times$48 mm$^2$. The area of each pixel is $5.8 \times 5.8$ mm$^2$. 
Instead of reading directly the 64 output channels of the PSPM, which would require high resources in terms of data storage and analysis, we connect them as inputs of an integrated circuit AMIC2GR \cite{herrero10, conde14} as shown in  Fig. \ref{fig:detector}. This front-end electronics has been developed for medical applications \cite{ros12}. It consists of 16 buffers that deliver copies of the 64 inputs to 8 computation blocks, each of which outputs a linear combination of the copies called moments.  

As described in Ref.\cite{ros12}, the moments are constructed according to 
$$ M_k = \sum_{i,j=0}^{i,j<8} c_{ijk} J_{ij},$$
where the index $k=0,...7$ runs over the moments, $J_{ij}$ is the copy of the input electric current of the PSPM pixel in row $i$ and column $j$ and $c_{ijk}$ are weighting coefficients. The latter are programmed using an I2C communication port integrated in the AMIC2GR circuit. For each moment $M_k$, the coefficients form a matrix of $8 \times 8$ entries that can be conveniently programmed in order to extract the relevant position or energy information, as described below. An important advantage of this readout approach is the fact that it is scalable \cite{herrero10}. This means that by coupling many detection systems together, a detector with a solid angle close to $4 \pi$ but still a relatively small number of output channels becomes feasible. This seems convenient towards a future implementation of a complex system, such as the $4 \pi$ i-TED described in Ref.\cite{iTED}. The individual currents $J_{ij}$ are series of electrical pulses, each of which corresponds to an event, and typically lasts around 50 ns. When using excessively high coefficients $c_{ijk}$, the resulting moments can give rise to saturated signals that are unexploitable. Thus, the magnitude of the coefficients needs to be adjusted for each specific measurement acccording to the energy range of interest.

\begin{figure}[!htbp]
\centering
\includegraphics[width=0.75\columnwidth]{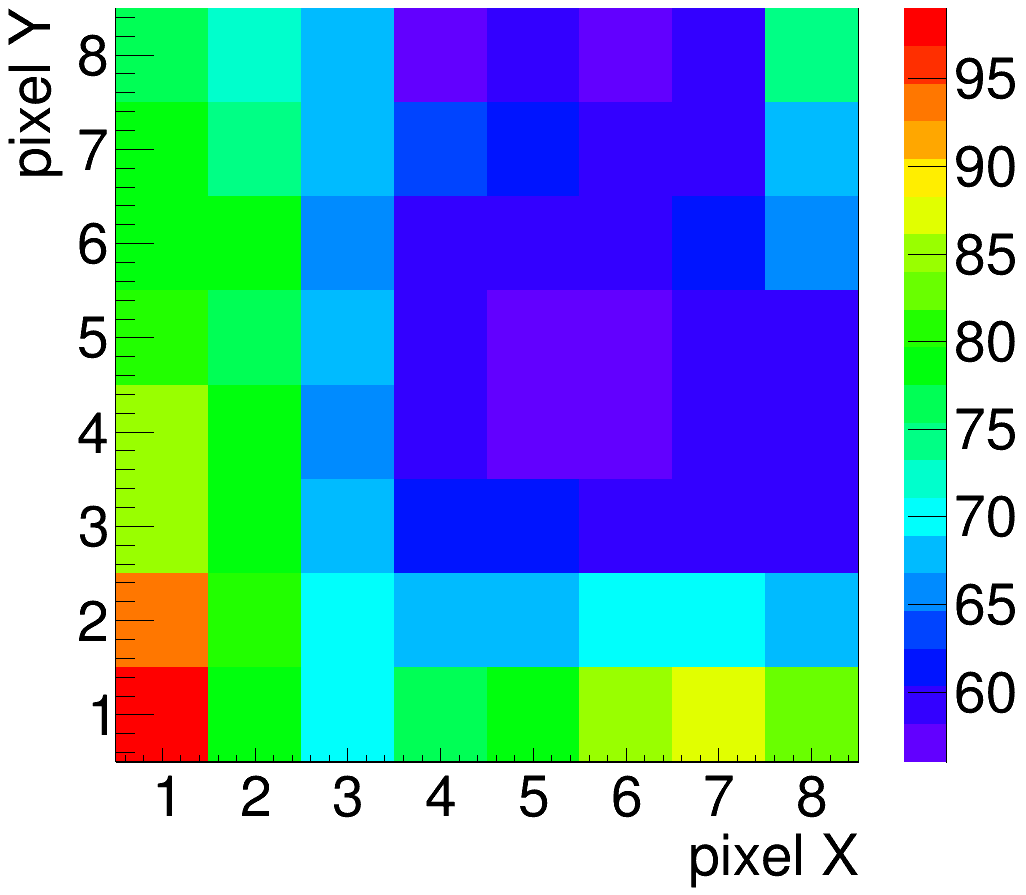} 
\caption{Visualization of the matrix of coefficients for $M_0$.} \label{fig:gaincorrectionmatrix}
\end{figure}

\begin{figure}[!htbp]
\centering
\includegraphics[width=0.49\columnwidth]{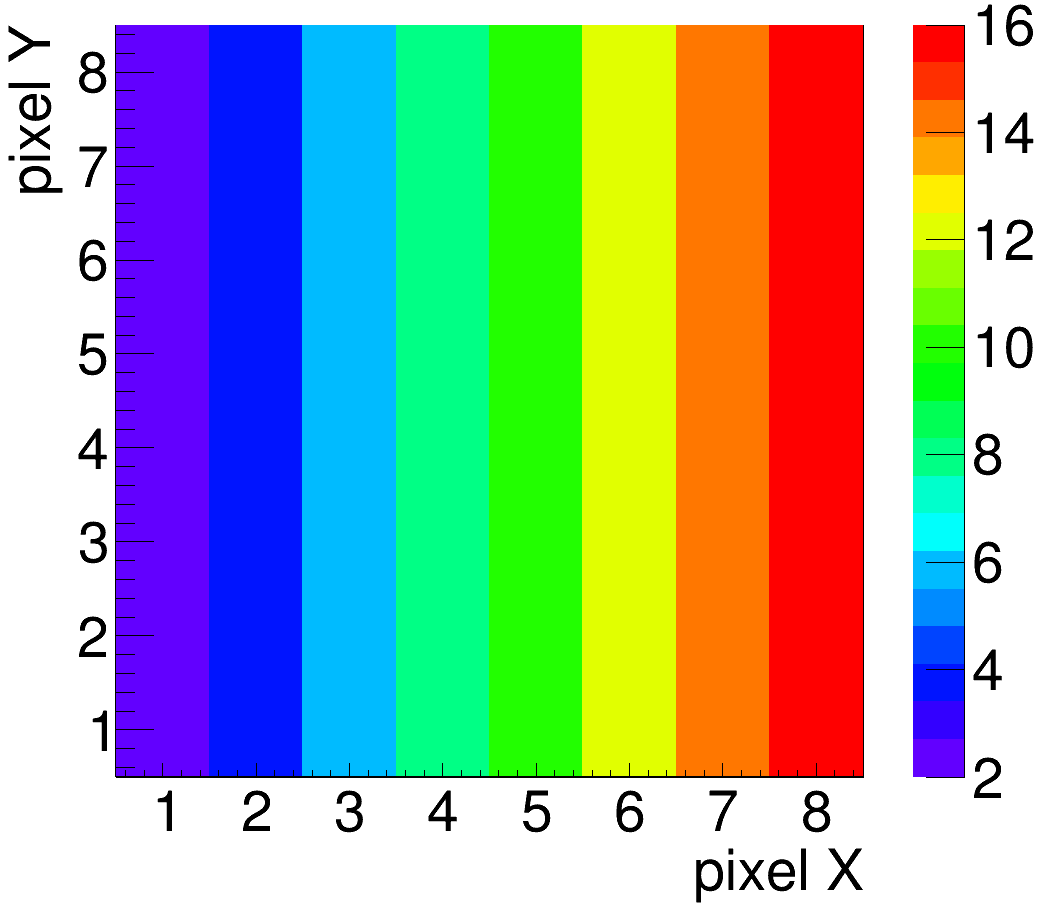} 
\includegraphics[width=0.49\columnwidth]{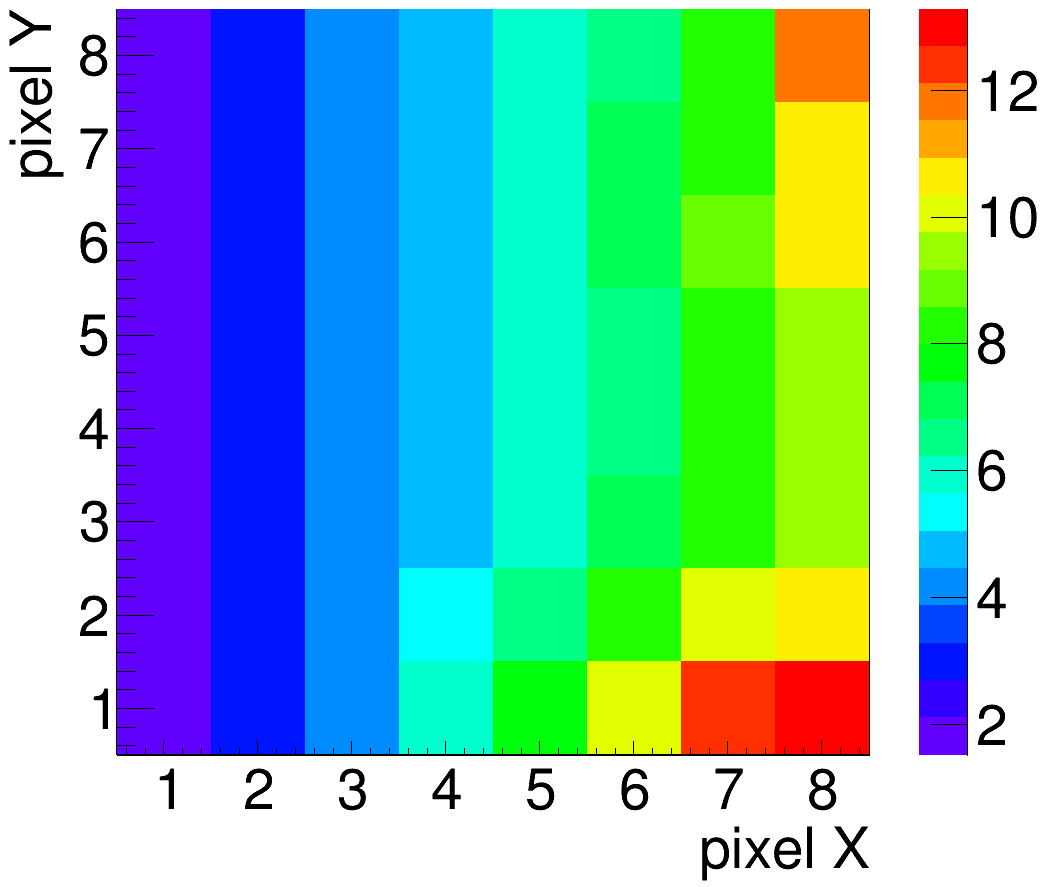} 
\includegraphics[width=0.49\columnwidth]{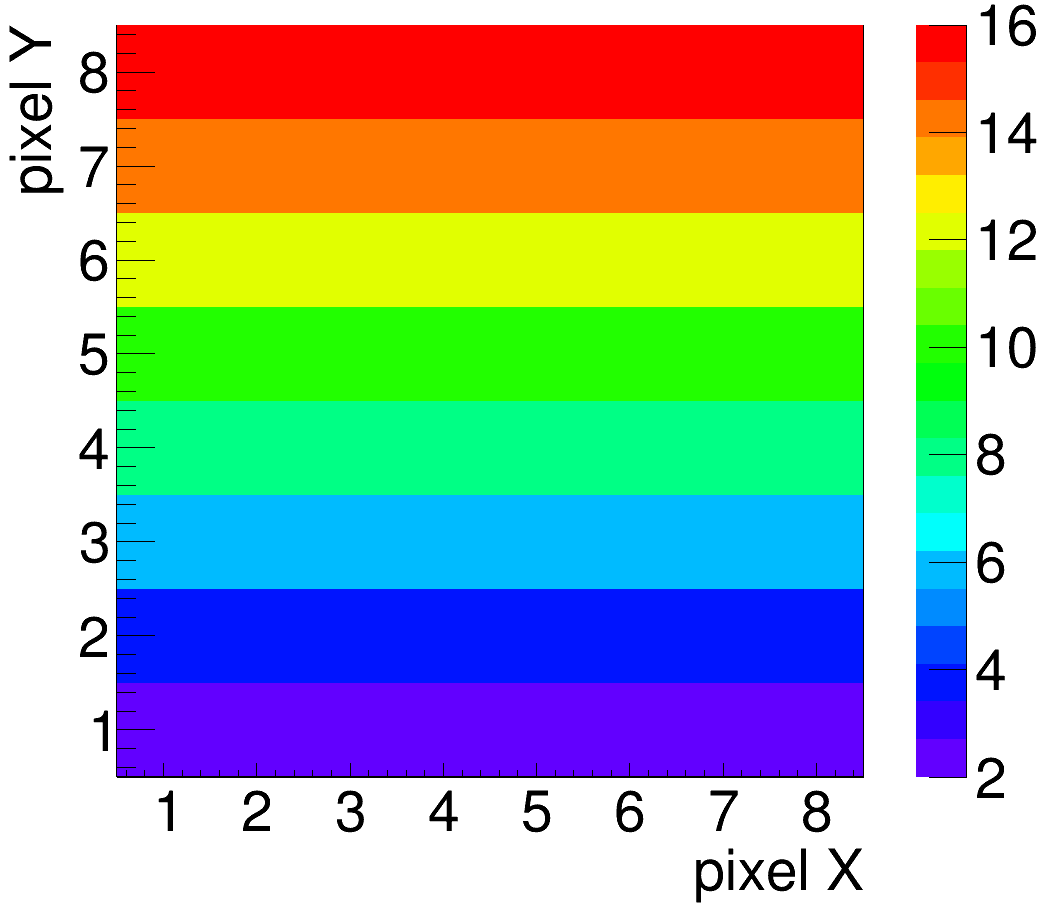} 
\includegraphics[width=0.49\columnwidth]{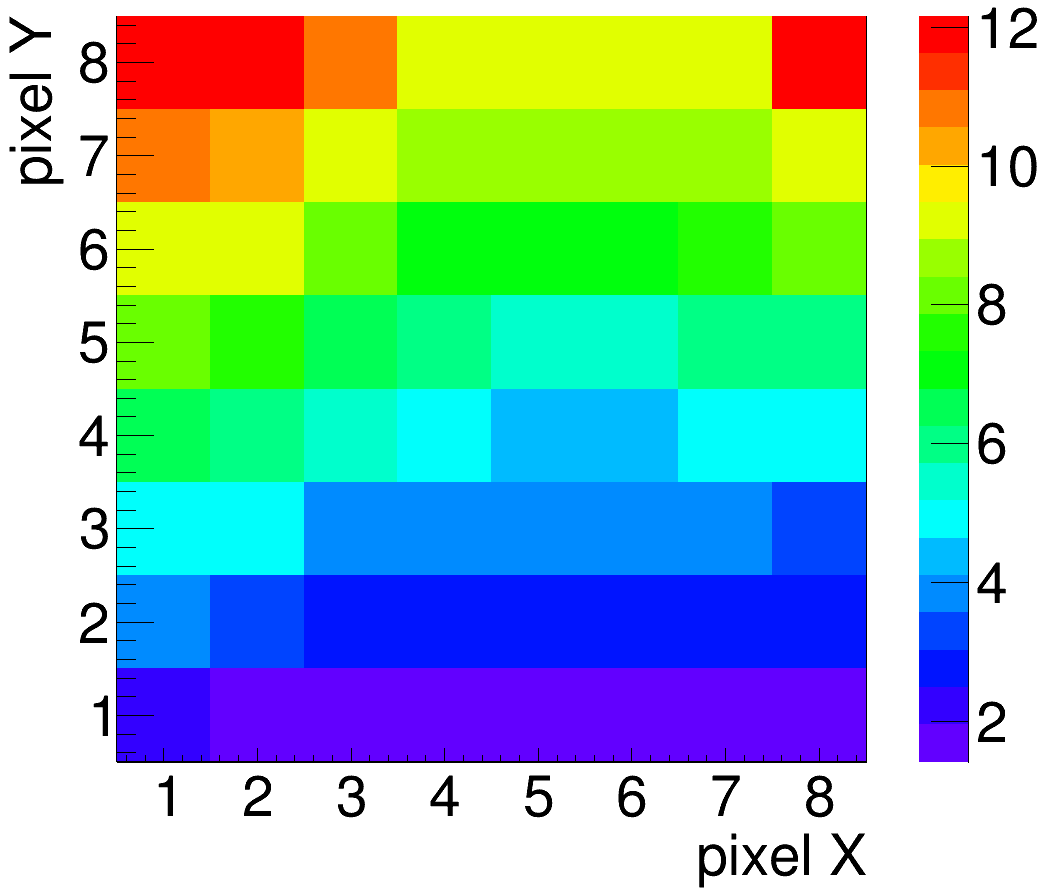} 
\caption{Visualization of the coefficients for M1 (top) and M2 (bottom). The  original horizontal gradient for M1 (top left) is corrected by the pixel gain inhomogeneities, resulting in the top-right matrix. The same is done for the M2 matrix (bottom panels).} \label{fig:matrixforxy}
\end{figure}

The methodology\cite{conde14,ros12} commonly employed with the AMIC2GR frontend electronics to reconstruct positions is similar to the working principle of an Anger camera~\cite{anger}, thus requiring four different coefficient matrices to recover the position of detection $(x,y)$ of an event. The approach followed in the present work is simpler and based on the use of only three moments ($k=0,1,2$): one to recover the photon energy $E$, and two to recover the position of detection $(x,y)$ on the PSPM surface. To recover energy $E$, we implement for the first moment $M_0$ a uniform matrix (all coefficients equal) corrected by a matrix $M$ that counterbalances the inhomogeneities in gain response between the pixels of the photocathode provided by the manufacturer. The final matrix is shown in Fig.\ref{fig:gaincorrectionmatrix}. To recover the position $x$, a positive horizontal gradient is implemented for $M_1$, resulting in an amplification of an energy deposition proportional to the position of detection along the $x$ axis. Ideally, to obtain the best position resolution, the gradient should have the highest slope possible. Nevertheless, the risk of saturation constrains its value. The linear coefficient gradients are corrected also by the gain inhomogeneities of the pixels $M$. In Fig. \ref{fig:matrixforxy}, we show the initial gradient (top-left) and corrected gradient (top-right) for $M_1$. By analogy, for position $y$, the same is repeated but replacing the horizontal gradient by a vertical one. Initial and corrected gradients for $M_2$ are shown in Fig. \ref{fig:matrixforxy} (bottom left and right).

An event on the sensitive volume of the gamma detector translates into three pulses, one in each moment, at the same time.  Pulses belonging to a same event are therefore identified through the time coordinate $t$. From the amplitudes $A_k$ of the pulses we build three quantities ($A_0, A_1/A_0, A_2/A_0$), from which we extract the useful information about the event ($E,x,y$) by supposing linear relations between these quantities and calibrating using reference points. Note that for $x$ and $y$ the pulse amplitude in the corresponding moments, $A_1$ and $A_2$ is divided by the pulse amplitude in the energy moment $A_0$, in order to obtain a quantity that does not depend on the energy of the event. 

The coefficients remain unchanged at all times throughout this work, as any change would affect the quality of the signal and modify the calibration in energy and position. The quality of the signal resulting from our choice of coefficients is checked by loading them into the AMIC2GR and placing a collimated source ($\varnothing  = 1$ mm) onto the surface of the detector, at the highest $(x,y)$ position (region with highest coefficients), and making sure that the moments are well below the saturation region. 

The use of a greater number of moments (we have 5 left) would provide a better performance of the system, but for this work we prefer to use a small number of channels that can be easily integrated in the digital acquisition system~\cite{daq} of n\_TOF (see Sec \ref{sec:measurementsatntof}). The voltage supplied to the PSPM is $-1000$~V, and the AMIC2GR requires a low-voltage supply of $6,12,-6$~V for built-in preamplifier stages and device operation.

\section{Detector setup and characterization}\label{sec:psd characterization}
The detection system is assembled and characterized at the Gamma-Ray and Neutron Spectroscopy Laboratory of IFIC for its posterior implementation at n\_TOF (see Sec.~\ref{sec:measurementsatntof}). The low voltage supply for the AMIC2GR is taken from a standard NIM crate and the high voltage for the photomultiplier was supplied by a TENNELEC (TC952) module. A self-triggered data acquisition system~\cite{agramunt13} is employed to digitize the electrical pulses corresponding to the three moments (M$_0$, M$_1$ and M$_2$) and extract the relevant information, namely the pulse-height and time. This acquisition system is based on SIS3316~\cite{sis3316} modules from Struck, which are operated at a sampling rate of 125~MHz and 14~bit of ADC resolution. The data is stored in ROOT\cite{root} trees and analyzed by implementing \texttt{C/C++} analysis routines. 

\subsection{Energy calibration and resolution}
\begin{figure}[!htbp]
\centering
\includegraphics[width=85mm]{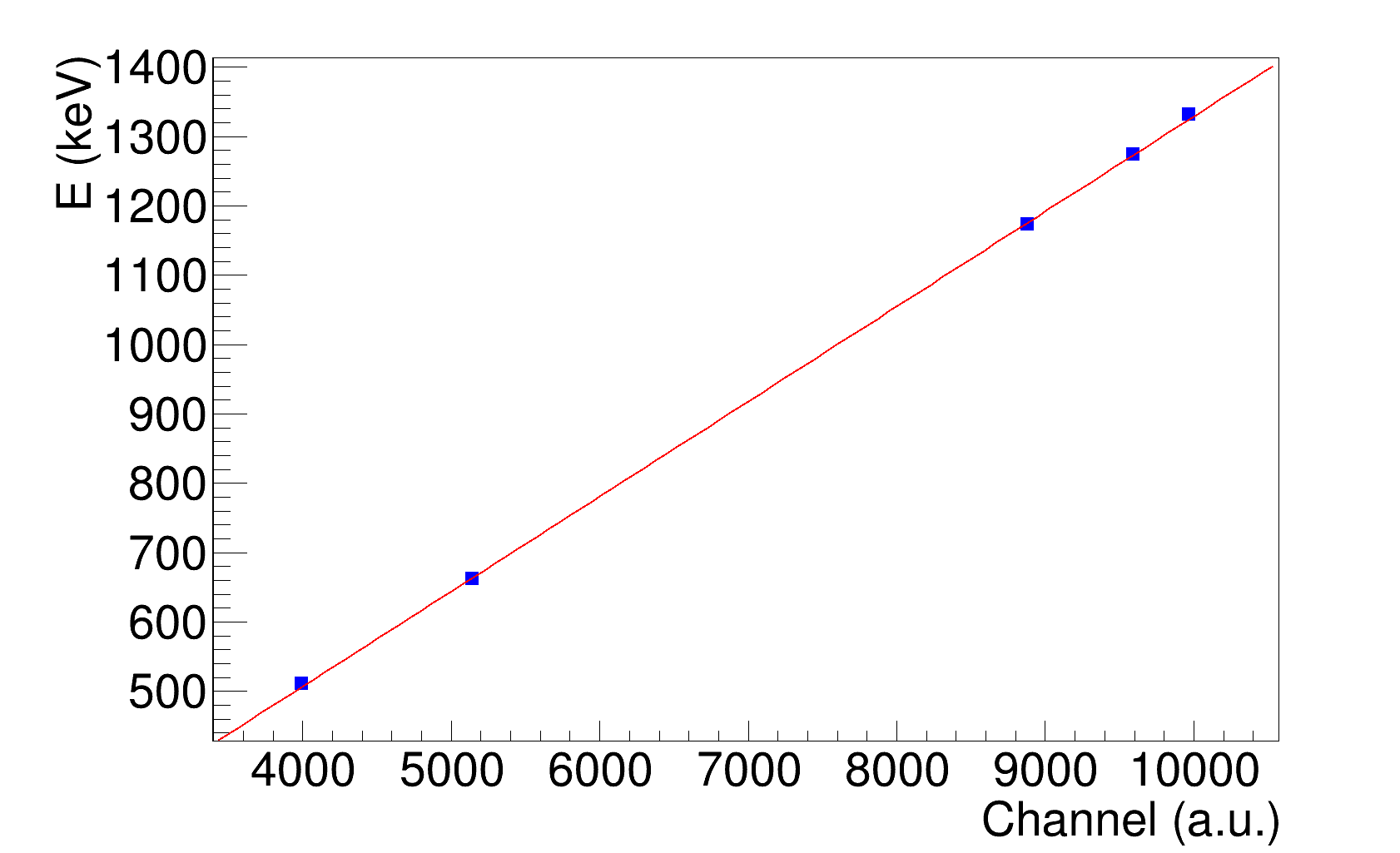} 
\caption{Energy calibration using $^{22}$Na, $^{60}$Co and $^{137}$Cs reference souces.} \label{fig:calibrationenergetique}
\end{figure}

An energy calibration is performed by using three different radioactive sources of $^{22}$Na, $^{60}$Co and $^{137}$Cs placed in turns at a distance of $5$ cm to ensure full illumination of the detector. As shown in Fig. \ref{fig:calibrationenergetique} the calibration shows a good linearity over the energy range up to 1.3~MeV. The energy resolution obtained at $662 $ keV is of 7.6\%~\textsc{fwhm}. 

\subsection{Spatial linearity} \label{ssec:reponse}
\begin{figure}[!htbp]
\centering
\includegraphics[width=80mm]{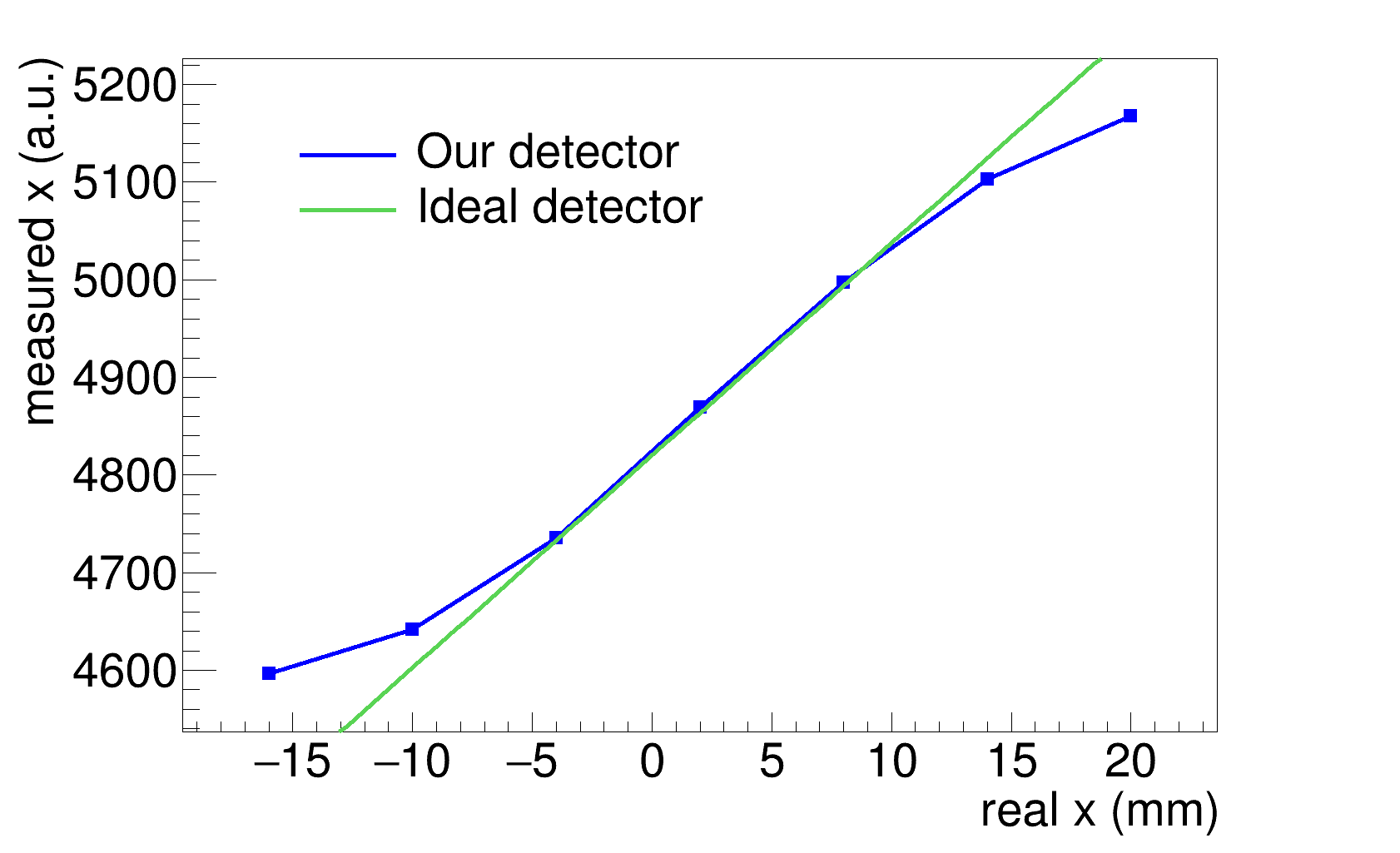} 
\caption{Uncorrected (raw) response of our detector and an ideal detector to displacements in $x$.}\label{fig:responsex}
\end{figure}

\begin{figure}[!htbp]
\centering
\includegraphics[width=80mm]{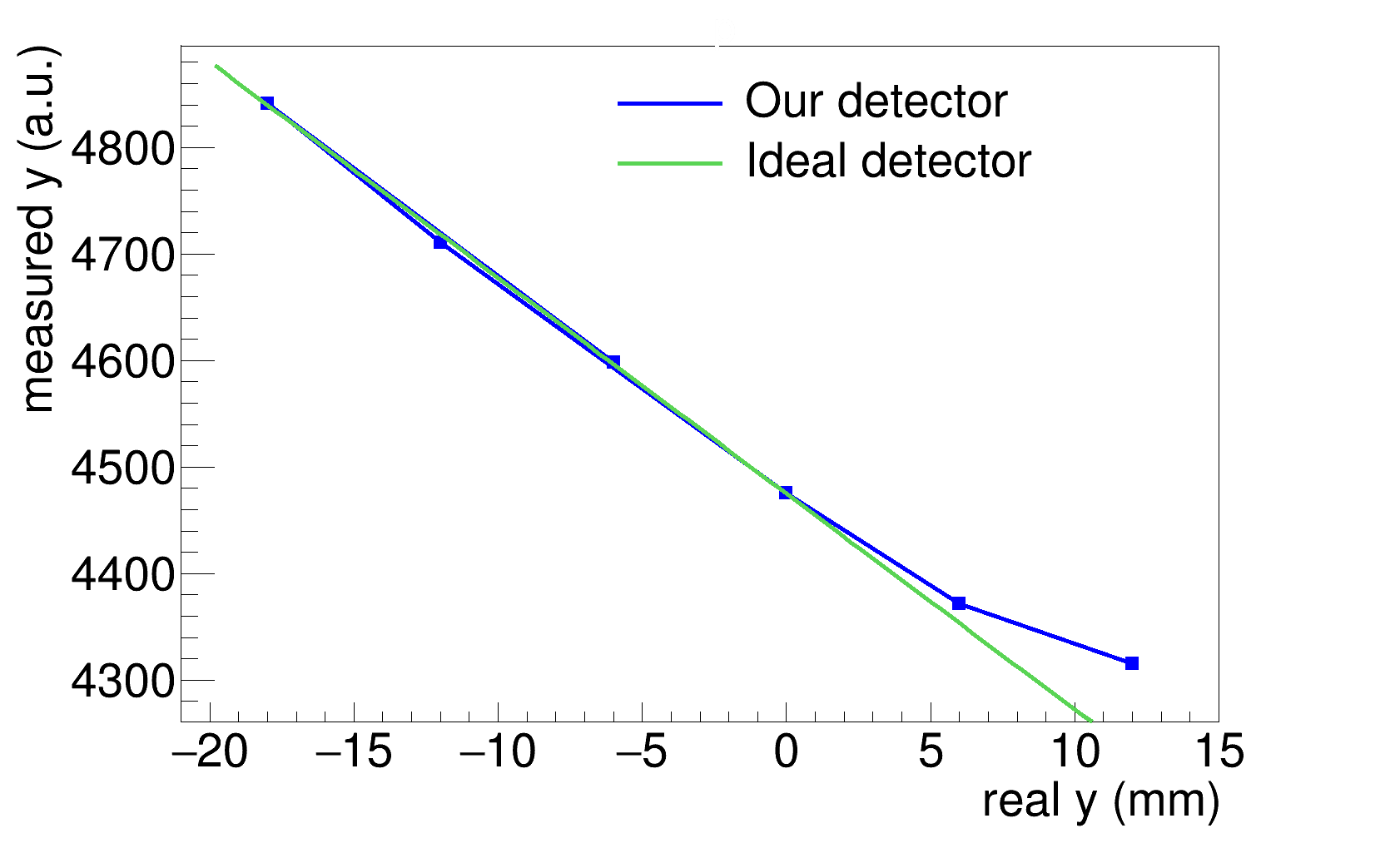}
\caption{Uncorrected (raw) response of our detector and an ideal detector to displacements in $y$.} \label{fig:responsey}
\end{figure}

Fig. \ref{fig:responsex} shows the correspondence between the real position $x$ on the surface of the detector of a collimated radioactive source of $^{137}$Cs  (290 kBq, $\varnothing_{col}  = 1$ mm) and the position measured by our detector. The data is obtained by using the scanning system described in Sec. \ref{ssec:reconstruction}. In these graphs, the compression due to border effects is visible. The same plot is obtained for the response in $y$ (Fig. \ref{fig:responsey}). An ideal detector (green line in Figs. \ref{fig:responsex} and \ref{fig:responsey}) has a linear response even for sources on the edges of its surface.  We determine that the range in which the response is linear is approximately $x \in [-10,15]$ mm and $y \in [-20,5]$ mm, in other words, the linearity region of the photocathode is estimated to be $25 \times 25$~mm$^2$. However, the useful field of view is a larger region, owing to the fact that the detector response can still be exploited before the linearity reaches saturation, by means of position reconstruction algorithms. This is shown in the next section.

%Encontrar el rango optimo. 
\subsection{Position reconstruction} \label{ssec:reconstruction}

\begin{figure}[!htbp]
\centering
\includegraphics[width=\columnwidth]{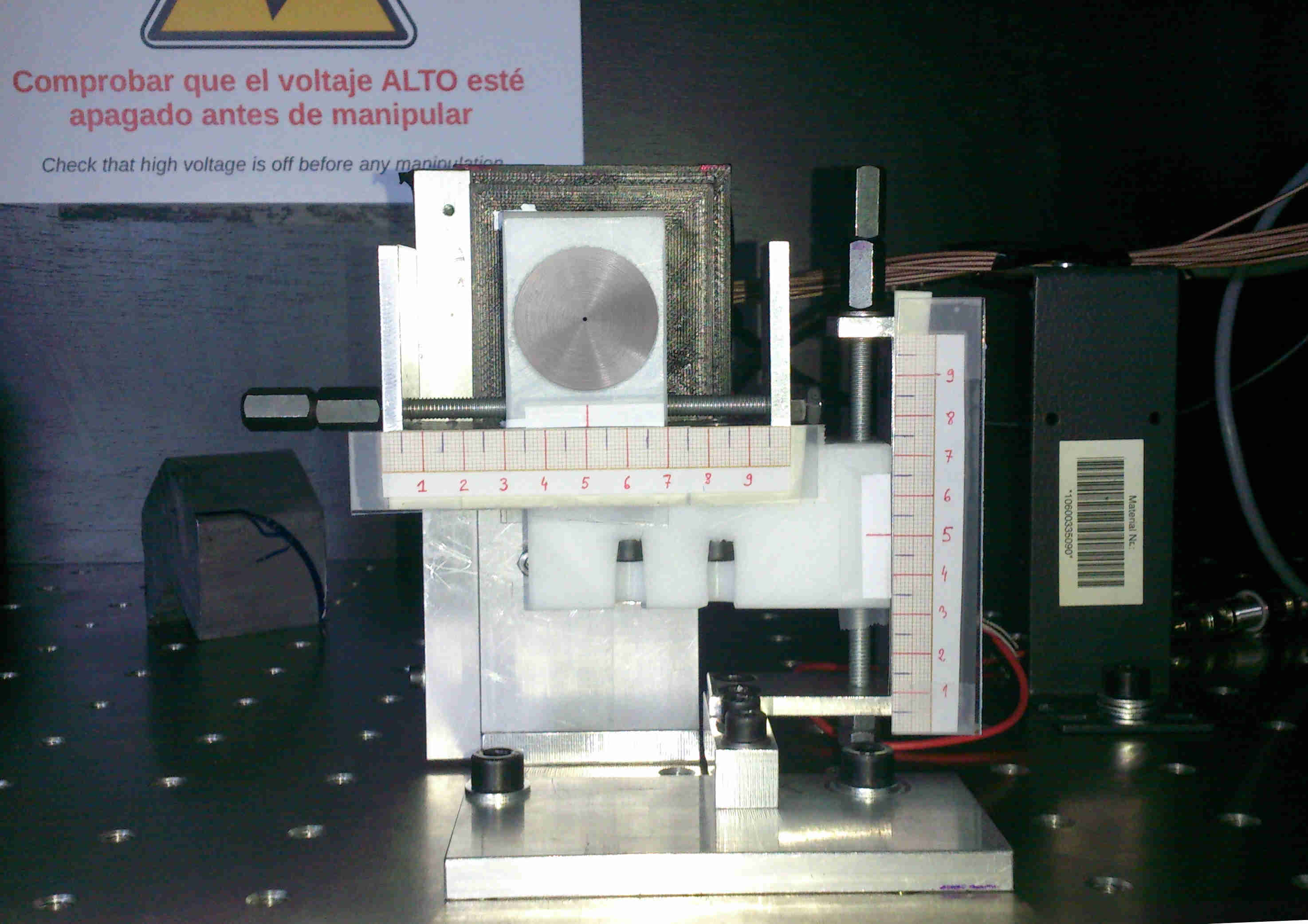} 
\caption{Scanning system with a 1~mm hole tungsten-collimator used to perform the spatial calibration of our detector.} \label{fig:setuplinearity}
\end{figure}

We now test the spatial capabilities of our detector in more detail. This is done by using the scanning system shown in Fig. \ref{fig:setuplinearity}. The position of the detector is kept fixed by placing it on a stand, and a collimated $^{137}$Cs source (290~kBq, $\varnothing_{col}  = 1$ mm) is displaced along the plane of the detector surface with a sub-millimetric precision by intervals of 6~mm horizontally and vertically. The moments are acquired for 64 positions of the source, corresponding to those of a 8 $\times$ 8 spatial grid, centered on the detector. The duration of each acquisition run is about $600$~s. For each \textit{real} position ($x_i, y_j$) of the source ($i,j=1,\ldots 8$), the mean value  of the pulse-height spectrum in the M1 momentum (counts vs. $\frac{A_1}{A_0}$) and the mean value of the M2 spectrum (counts vs. $\frac{A_2}{A_0}$) is determined. They represent the \textit{measured} position $x'_i, y'_j$. Following the methodology described in Ref.~\cite{berta} we define two functions, $f_1 = f_1 (x',y')$ and $f_2 = f_2 (x',y')$ given by
$$ f_1 (x',y') = \sum_{i=0}^M E_i \cdot x'^{a_i} \cdot y'^{b_i}$$

$$ f_2 (x',y') = \sum_{i=0}^N F_i \cdot x'^{c_i} \cdot y'^{d_i}$$

and find the values of all the parameters $\{a_i, b_i, c_i, d_i, E_i, F_i \}$ such that
\begin{equation}
(x_{R} \equiv f_1 (x',y') ~~, ~~ y_{R} \equiv f_2 (x',y'))
\end{equation}
is the best fit of the set of data points $\{ (x_i,y_j), i,j=1,\ldots 8\}$. This is done in ROOT by using the \texttt{TMultiDimFit} class.  Once the polynomial form is known, we can plot the reconstructed positions, $\{(x_{R,i}, y_{R,i}) \}$. The closer they are to the real positions, the higher the linearity, the field of view and the overall spatial performance.

\begin{figure}[!htbp]
\centering
\includegraphics[width=0.8\columnwidth]{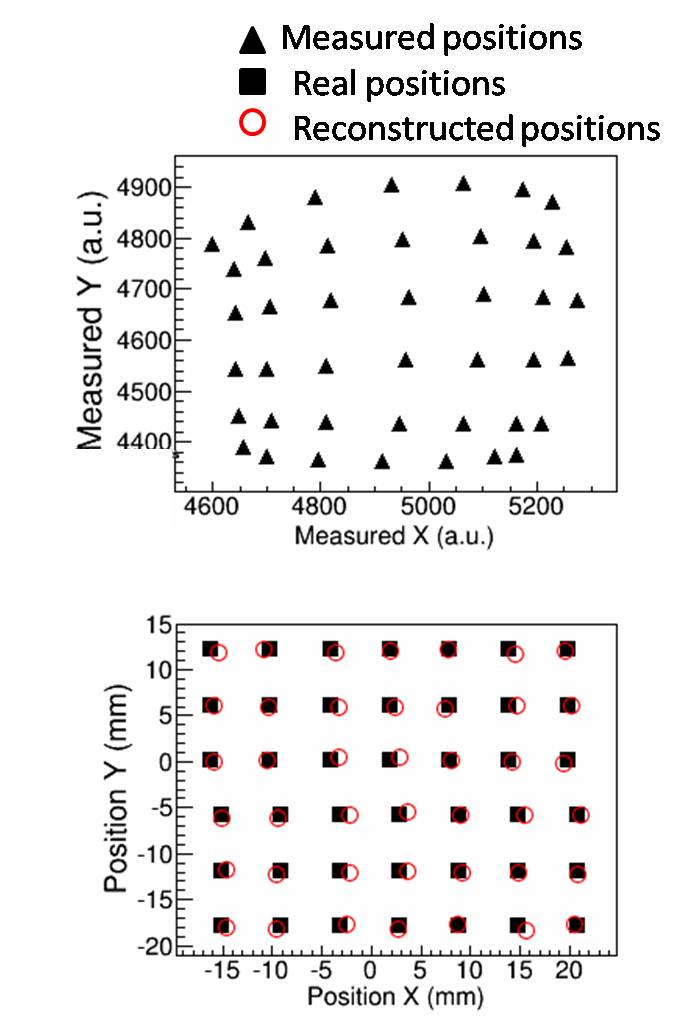} 
\caption{Grid of raw measured positions (top) and calibrated positions (bottom). Bold square symbols in the bottom panel show the grid of real positions used as reference.} \label{fig:reconstruction}
\end{figure}

The results of the position reconstruction are compiled in Fig. \ref{fig:reconstruction}. The top graph represents the measured positions for each real position. Some positions are discarded for being too close to the edge leading to strong distortion due to border effects \cite{berta}. The bottom graph in Fig. \ref{fig:reconstruction} compares the reconstructed positions with the real ones. The value of the average deviation was of 1~mm~\textsc{fwhm}. Since the reconstruction is successful, $f_1$ and $f_2$ can be used to calibrate spatially the detector over a useful field of view of about $35 \times 35$ mm$^2$.

\subsection{Collimator}
\begin{figure}[!htbp]
\centering
\includegraphics[width=36mm]{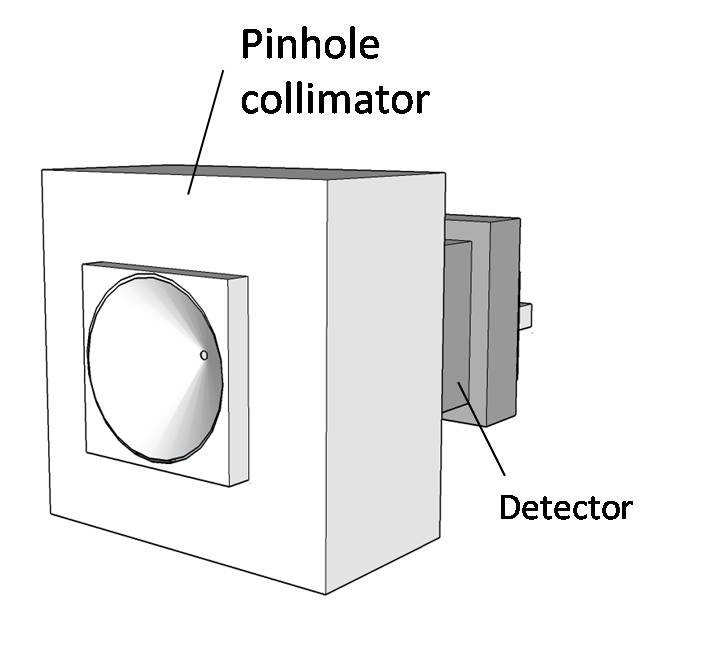}
\includegraphics[width=36mm]{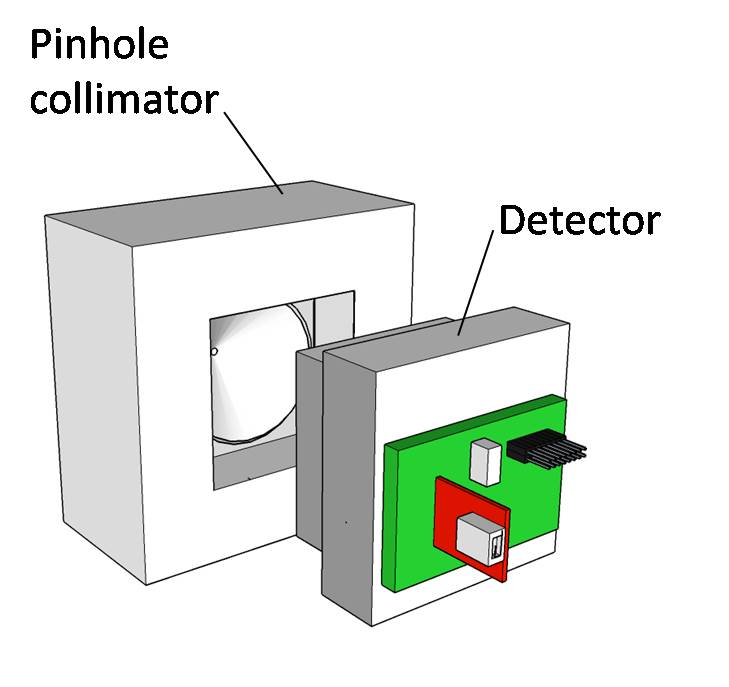}
\caption{Schematic view of the pinhole lead collimator and the position-sensitive detection system.} \label{fig:camera}
\end{figure}

Finally, to build a $\gamma$-ray imager, we couple mechanically the position-sensitive detector described above to a pinhole collimator. The collimator is a block of lead carved into two inverted cones, as shown in Fig \ref{fig:camera}. It is varnished with a layer of paint to protect from the toxicity of lead. The pinhole aperture is of 3~mm in diameter and the distance between the pinhole and the front surface of the scintillating crystal is 4~cm.

The idea behind the geometry of the collimator is that an event detected on a point $P$ of the detector surface can only come from the direction defined by this point $P$ and the pinhole, since lead absorbs gamma radiation. The collimator was designed by means of \textsc{Geant4}~\cite{geant4} Monte-Carlo simulations in such a way that its shielding efficiency is around 85\% for energies of 1.3~MeV. There is therefore a one-to-one correspondence between the direction (or angle) of incidence of a $\gamma$-ray and the point of impact on the surface of the detector. Since the effective photocathode area is $48\times 48$ mm, we estimate the total angular field to be $\theta_{eff} \approx 24^\circ$. The angular field with linear response is $\theta_{lin} \approx 17^\circ$. Another important fact is that the image is inverted. This effect will be directly corrected by software. 
The performance of the gamma camera composed of this collimator in addition to the position-sensitive detector will be illustrated in Sec.~\ref{sec:analyse} on the basis of dedicated calibration measurements using radioactive sources.

\section{Measurements at n\_TOF} \label{sec:measurementsatntof}

In this section, we describe the part of the work carried out at CERN n\_TOF. We first introduce the reader to the n\_TOF facility and then detail the measurements carried out with our gamma imager at the measuring station EAR1.

\subsection{Introduction to the facility} \label{ssec:facility}
At the n\_TOF experiment at CERN \cite{guerrero13}, a very intense  pulsed proton beam ($7 \times 10^{12}$~protons/pulse) impinges onto a lead block, generating about $300$ neutrons by means of spallation reactions for each incident proton. The spallation also produces charged particles and gamma radiation, which are significantly attenuated from the beam by use of a combination of collimators and magnets. The neutrons impinge onto the targets of interest in two different experimental areas, EAR1 and EAR2, located respectively at $185$ and $20$ m from the spallation source. In both cases, the energy of an incident neutron is recovered by measuring the time-of-flight (TOF). This standard technique is based on the fact that given a beam of neutrons created at the same time, there is a direct relationship between the kinetic energy of a neutron and the time it takes to reach the target. The gamma flash produced when the pulse of protons impinges onto the spallation source is taken as the time origin $t_0$. Since EAR2 is much closer to the spallation source than EAR1, it provides a higher neutron flux. In turn, the longer tunnel to EAR1 offers a correspondingly better energy resolution. The two areas are therefore complementary. 

\subsection{Experimental setup} \label{ssec:setupatntof}
\begin{figure}[!htbp]
\centering
\includegraphics[width=\columnwidth]{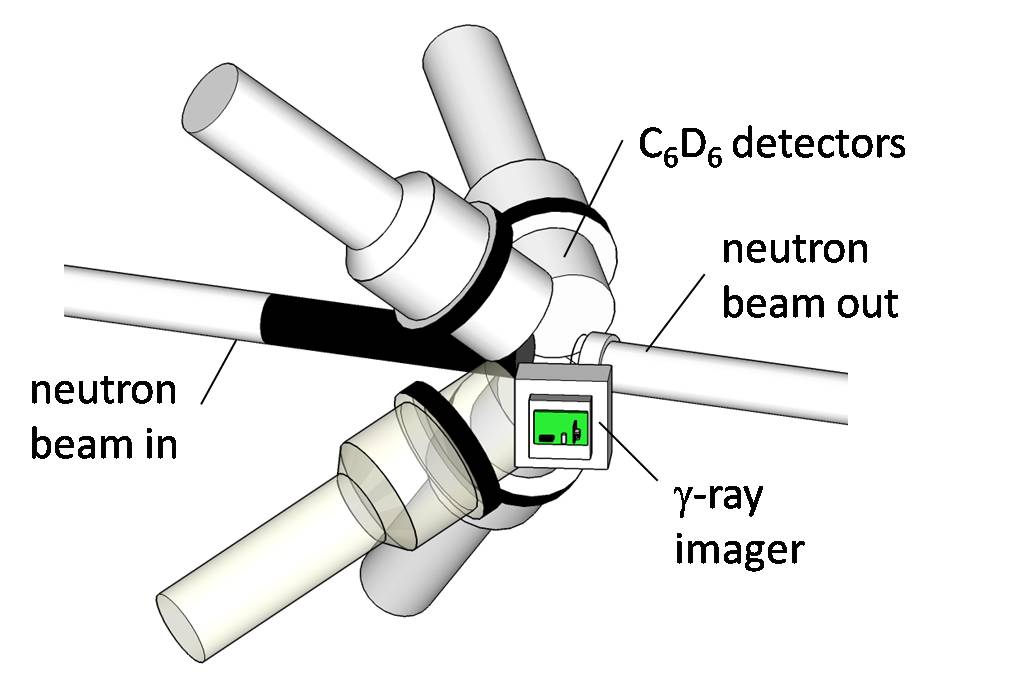} 
\caption{Schematic view of the experimental set-up displaying the four C$_6$D$_6$ detectors and the $\gamma$-ray imager. } \label{fig:oreille1}
\end{figure}

\begin{figure}[!htbp]
\centering
\includegraphics[width=\columnwidth]{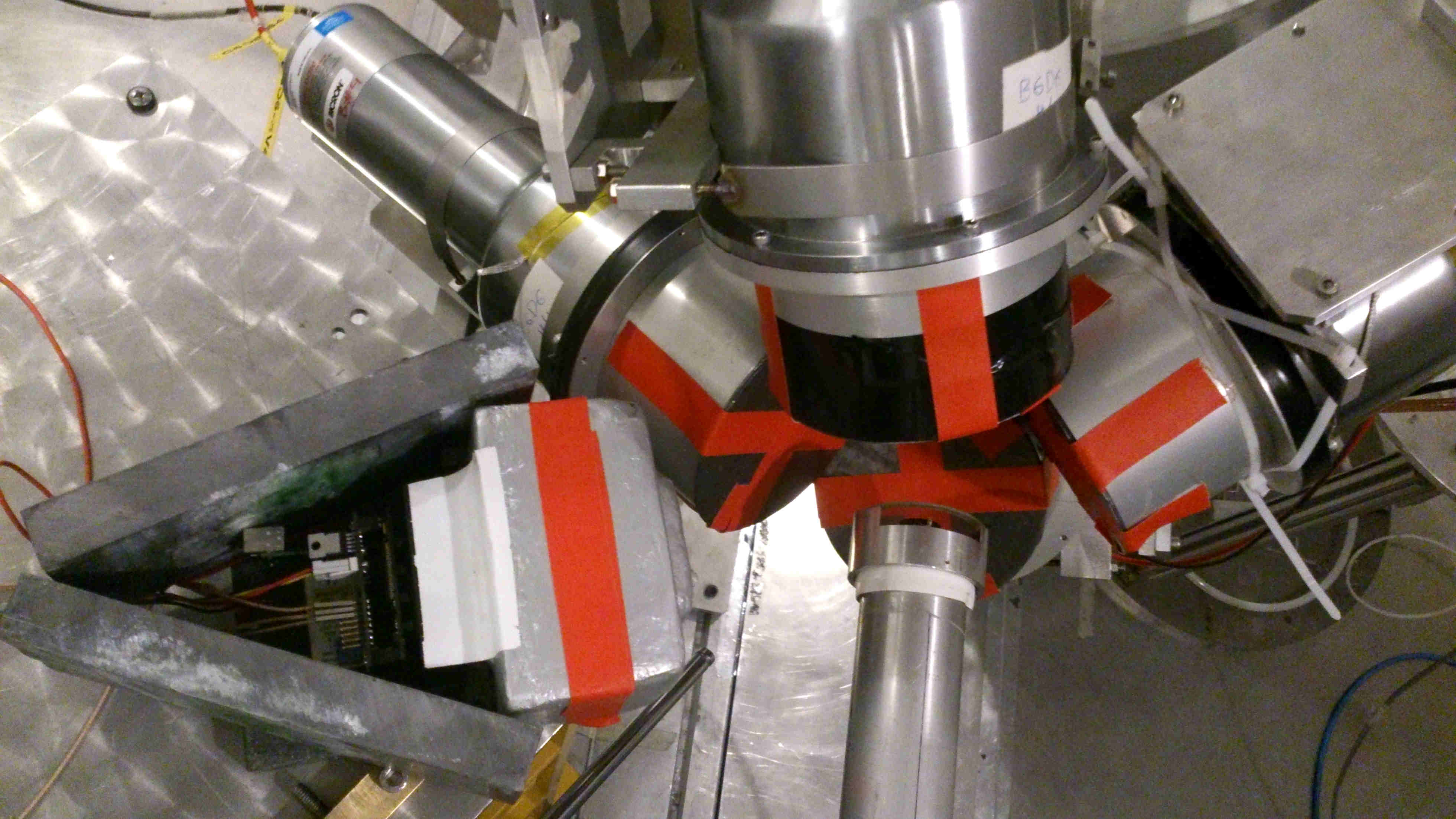} 
\caption{Setup as seen from above with four C$_6$D$_6$ detectors and the gamma-camera. The visible pipe is the neutron exit beam.} \label{fig:vraiephoto}
\end{figure}

The present measurements are performed at the EAR1 experimental area. Our gamma imager is placed in the pre-existing experimental setup of four C$_6$D$_6$ detectors used for neutron capture measurements, as shown in Figs. \ref{fig:oreille1} and \ref{fig:vraiephoto}. The C$_6$D$_6$ detectors were prepared for the measurement of a radioactive sample and for that reason a shielding layer of lead with a thickness of 2~mm was attached to their front surface.

The prompt $\gamma$-rays generated by the neutron capture reactions are detected by four C$_6$D$_6$ liquid scintillation  detectors. These detectors are characterized by a very low neutron sensitivity, but do not have spatial sensitivity. By contrast, our gamma camera is capable of imaging but has a very high neutron sensitivity, due to its massive lead collimator and the bromine contained in the crystal, that has a quite high neutron capture cross section dominated by resonances. Our gamma imager is placed in the remaining space, as close as possible to the target, at a distance of $(15.3 \pm 0.2)$ cm from its center, and forming the smallest possible angle $(18 \pm 0.5)^{\circ}$  with respect to the target plane. This is done so that the target subtends the smallest solid angle and thus confine true capture $\gamma$-rays within a small region in the center of the linear field of view of our detector. Two blocks of lead are placed on the rear side of the imager to shield it from contaminant rays coming from the sides and from the back (see Fig.~\ref{fig:vraiephoto}). The low voltage supply for AMIC2GR is provided by a NIM crate, and the high voltage for the photomultiplier by a CAEN A1733 module.  The three moments from the AMIC2GR are connected to three channels of the n\_TOF digital  acquisition system \cite{daq}. Each channel is digitized with 12 bits at a sampling rate of $500$ Msamples/s using a full scale of 1~V. 

\subsection{Data processing}\label{ssec:daq}
The three moments of the AMIC2GR are read and recorded but the time $t$ and amplitudes $A_0, A_1$ and $A_2$ of the pulses are extracted by means of a customized pulse-shape algorithm developed at n\_TOF~\cite{Zugec15}. Only events for which all three amplitudes are simultaneously detected will be kept, since they correspond to events for which energy and position can be obtained simultaneously, which is a fundamental requirement for our later analysis. 
From this data different diagrams are built for data analysis and interpretation. The first one is an XY map of counts, that displays a 2D image of the number of events detected as a function of $A_1/A_0$ and $A_2/A_0$, which are quantities proportional to $x$ and $y$ positions of detection, respectively. The second one is the energy spectrum, i.e. counts as a function of $A_0$, which is proportional to $E$. Position and energy calibrations are carried out, as described in Sec.~\ref{sec:measurements}, in order to present these diagrams in physically relevant units. We empower our analysis by implementing the possibility of specifying spatial cuts in the XY map, and generate energy spectra where only events belonging to these regions appear. Cuts based on energy are also used. Finally, we build a spectrum where we represent the number of $\gamma$-rays detected as a function of the TOF or, equivalently, to neutron energy. This spectrum is generated with and without spatial cut. It is the spectrum of greatest physical relevance, since it compares a typical measurement at n\_TOF with and without spatial discrimination, and allows us to see how the signal-to-background ratio changes.

\subsection{Description of the measurements}\label{sec:measurements}
\begin{figure}[!htbp]
\centering
\includegraphics[width=0.85\columnwidth]{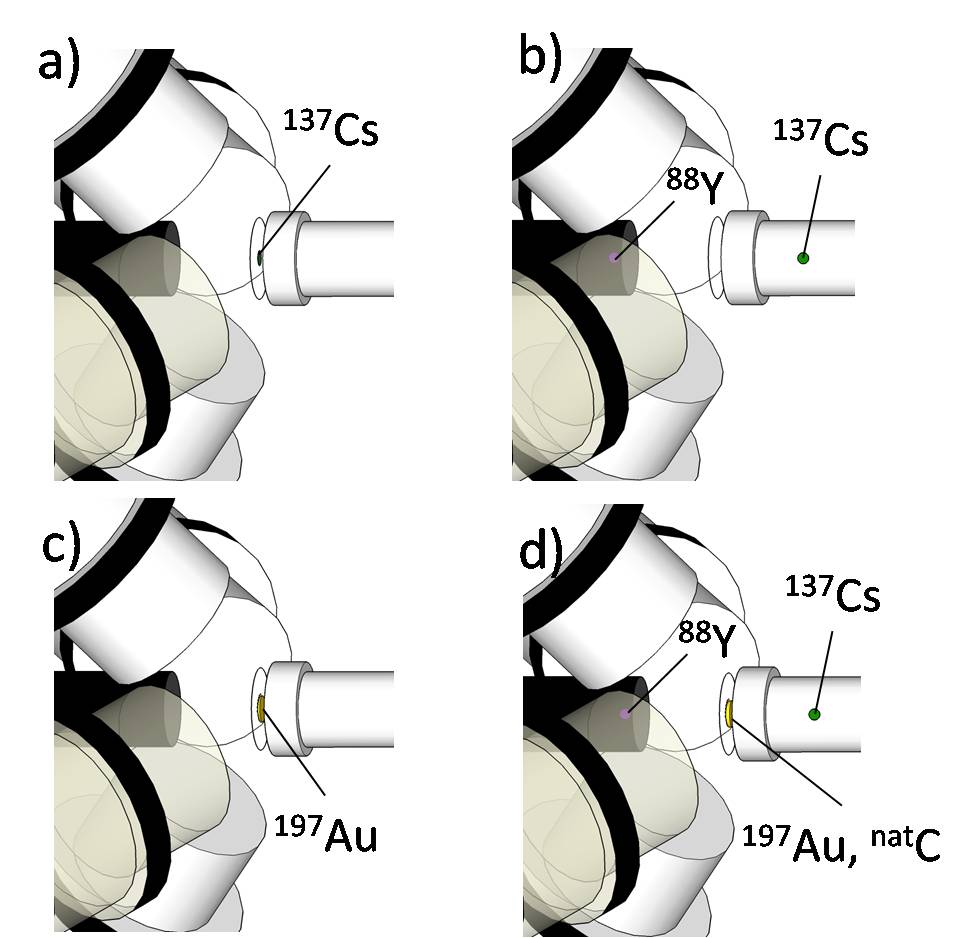}
\caption{The four measurements carried out at n\_TOF. (a) Calibrated $^{137}$Cs source in the center of the sample-position; (b) $^{88}$Y and $^{137}$Cs sources placed at $\pm$7~cm along the beam-pipe; (c) ($n,\gamma$) measurement using a $^{197}$Au sample and (d) ($n,\gamma$) measurement using a composite sample of gold and graphite together with the radioactive sources of (b). See text for details.} \label{fig:measurements}
\end{figure}

Four measurements are carried out. The setup configuration in each case is schematically shown in Fig. \ref{fig:measurements}. The two first measurements intend to provide a calibration in energy and position for our gamma camera. This is done by setting up two different configurations of radioactive sources and acquiring data with the neutron beam off. In (a), a $^{137}Cs$ source (activity: 417~kBq) is placed at the target location, for this is a crucial reference point. In (b) two sources, $^{88}Y$ ($377$ kBq) and $^{137}Cs$ (417~kBq), are placed along the beam pipe, respectively $(7.0 \pm 0.1)$ cm upstream and (7.0 $\pm$ 0.1)~cm downstream. 

In the two next measurements, a disk of $^{197}$Au 20~mm in diameter and 0.1~mm in thickness is placed at the target location. This stable isotope is chosen because its neutron capture cross section is high, thus allowing us to carry out the present measurements in a short time interval, and well-known, which is important to extract reliable conclusions. Using the neutron beam, the $^{197}$Au $(n,\gamma)$ reaction is measured in two situations. First in geometry (c) with only the $^{197}$Au sample. This measurement is a typical yield calibration measurement at n\_TOF and differs only by the use of the present $\gamma$-ray imager. It is relevant since we use it to prove that the S/B ratio in the TOF spectra can be enhanced by implementing spatial discrimination. We also want to know if this enhancement is possible in more severe background conditions. This is of interest in particular for $i)$ measurements of capture samples available in very small quantities, $ii)$ measurements of isotopes with a dominant neutron scattering channel, and $iii)$ measurements at n\_TOF EAR2 where the proximity of the experimental area to the spallation target implies a correspondingly higher background level. Thus, in setup (d), we try to reproduce a high background environment by increasing artificially the background of the case (c). This is done by adding the radioactive sources of measurement (b) as a localized source of background as well as by adding onto the gold disk another disk of same diameter made of natural carbon, with a thickness of 10~mm. Natural carbon has a very low neutron capture cross section, and a high neutron scattering cross section. As shown below, the result is a substantial increase of the overall background level.

\begin{figure}[!htbp]
\centering
\includegraphics[width=\columnwidth]{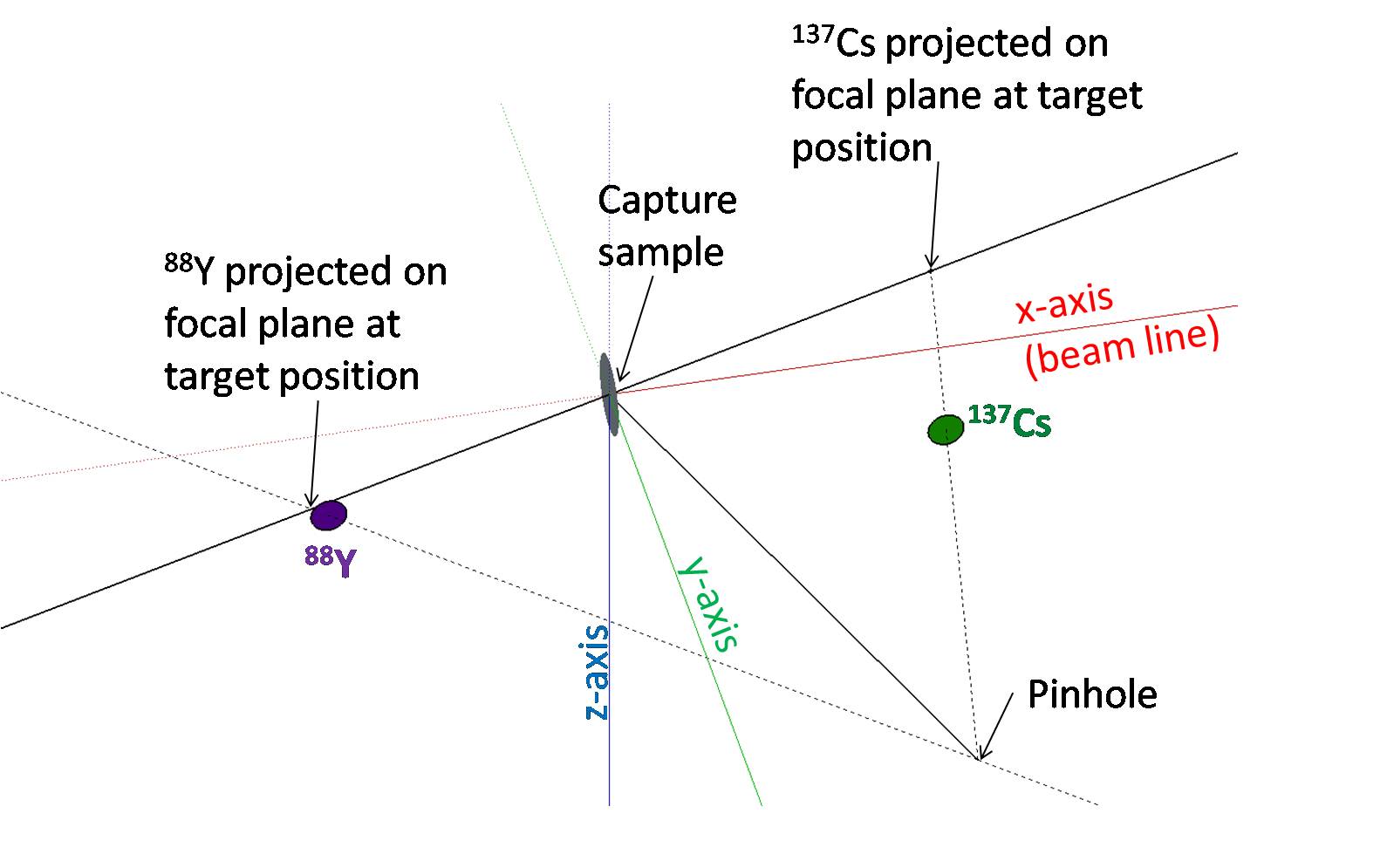} 
\caption{Projection of the sources on the focal plane at target position.} \label{fig:geometry}
\end{figure}

\section{Analysis and results} \label{sec:analyse}
First of all we carry out an in-situ spatial calibration of the $\gamma$-ray imager. Since it is sensitive to directions but not to the distance of sources, the real positions of objects in the setup was  referred to using their optical-projected position $(x,y)$ onto the focal plane of the gamma camera at sample position. An illustration of this projection is shown in Fig. \ref{fig:geometry} for the setup in Fig. \ref{fig:measurements}b . In these new coordinates, the sample location remains $(x,y)=(0,0)$ which is defined by the $^{137}$Cs measurement (a), and we find for measurement (b) the positions $^{137}$Cs $(x,y)=(85.0, 0)$ mm and $^{88}$Y $(x,y)=(-78.3,0)$ mm.  If we assume that the real shape of the collimator is very close to the desired geometry, the good linearity of the surface of the detector should translate into a good linearity in this focal plane. Although these points lie close to the boundaries of the efficient angular field, we neglect the border effects and calibrate our gamma camera spatially using these three points. For the $y$ coordinate the calibration is assumed to be the same, which is justified after the similar linearity response shown in Fig.~\ref{fig:responsex} and Fig.~\ref{fig:responsey}. For the energy calibration, the dominant peaks of $^{137}$Cs (662 keV) and $^{88}$Y (898,1836 keV) are identified in the energy spectra. 

\begin{figure}[!htbp]
  \centering
  \includegraphics{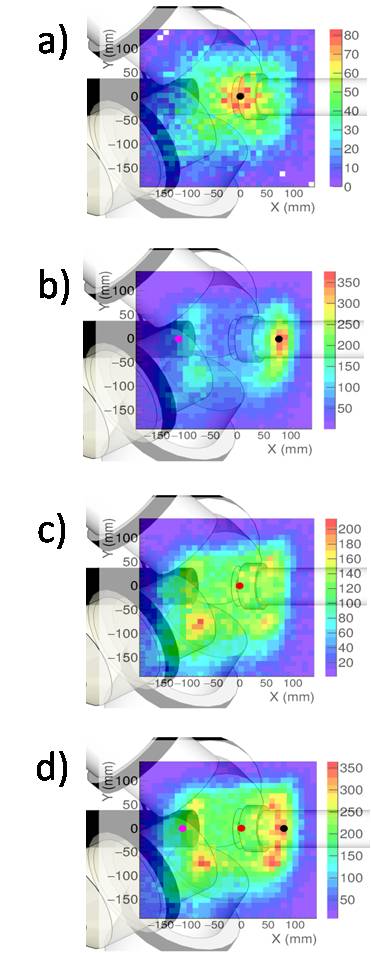}
  \caption{XY $\gamma$-images obtained for the four configurations shown in Fig.\ref{fig:measurements}. For the sake of clarity the geometry of each configuration is shown in the background. Solid symbols indicate the location of the radioactive sources and samples as depicted in Fig.\ref{fig:measurements}.} \label{fig:superpose}
\end{figure}

We first look at the XY map of counts obtained for each measurement. They are shown in Fig. \ref{fig:superpose}, superimposed onto the laboratory configuration. These \textit{gamma images} are the first of this kind to be done at n\_TOF and allow one to make a qualitative diagnostic of the different $\gamma$-ray and background sources contributing to the detector response. In Fig. \ref{fig:superpose}a, the $^{137}$Cs source translates into a well-defined zone of higher counts in the center of the image. In Fig. \ref{fig:superpose}b, the two radioactive sources appear as two distorted patterns on both sides of the image. Their deformed shape should be compared to the sides of the top graph of Fig. \ref{fig:reconstruction}. It is clearly a result of pin-cushion effects since, as mentioned earlier, these sources lie on the boundaries of the efficient angular field of view. The elongated shape along the y-axis may be ascribed to backscattering from the two shielding Pb-blocks placed behind the detector (see Fig.~\ref{fig:vraiephoto}). The left-region of the gamma image corresponding to the $^{88}$Y source has less counts because it is partially screened by the Pb shielding of the C$_6$D$_6$ detectors.  Fig. \ref{fig:superpose}c shows the $^{197}$Au ($n, \gamma)$ measurement. The sample does not appear as a region of higher counts due to strong background, and the regions of higher counts on both sides are also ascribed to backscattering by the Pb blocks that were placed on both sides behind the detector. Finally, in Fig. \ref{fig:superpose}d, we obtain simultaneously the patterns of Figs. \ref{fig:superpose}b and \ref{fig:superpose}c.

\begin{figure}[!htbp]
\centering
\includegraphics[width=\columnwidth]{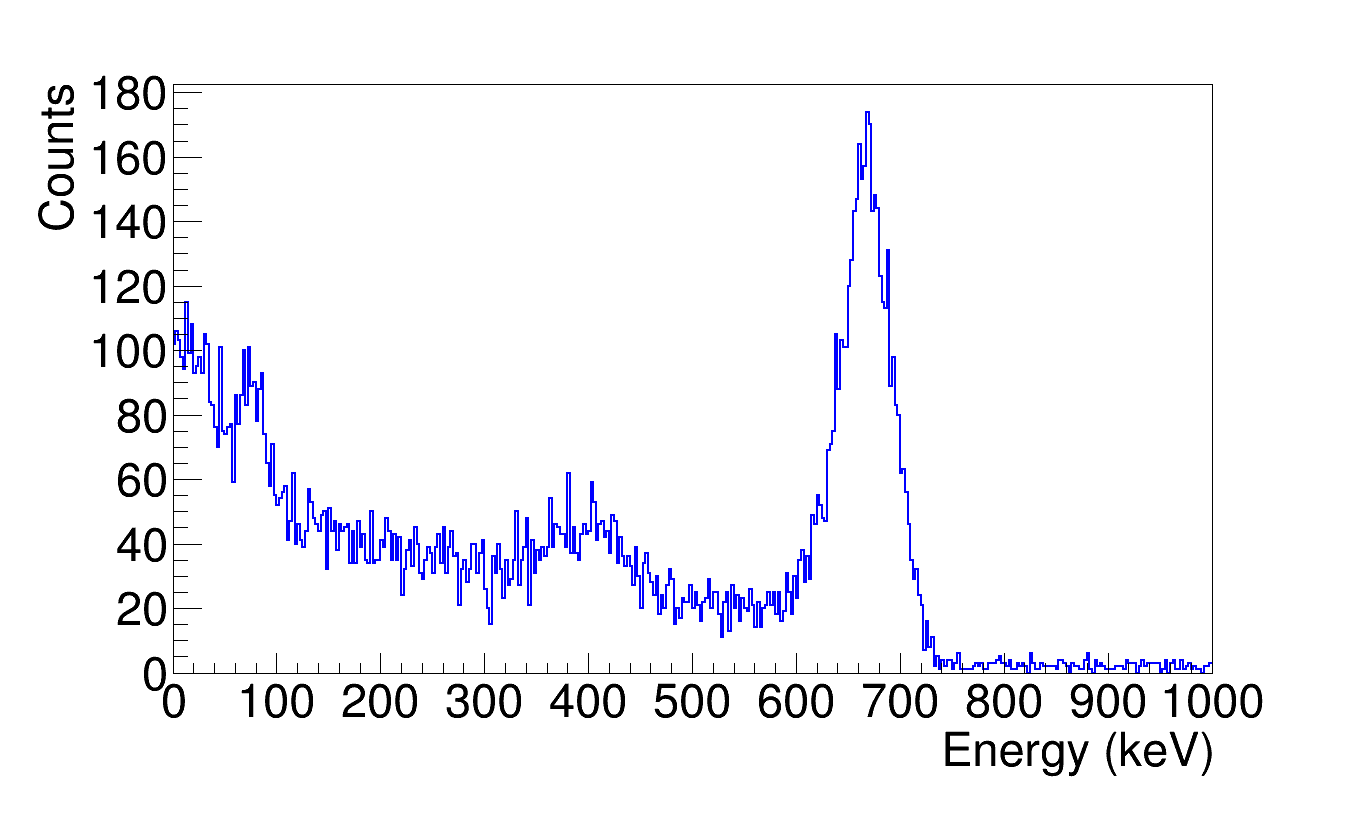}
\caption{Energy spectrum for measurement (a) where the $^{137}$Cs source is placed at sample location (Fig.\ref{fig:superpose}a)} \label{fig:energya}
\end{figure}

\begin{figure}[!htbp]
  \centering
  \includegraphics[width=\columnwidth]{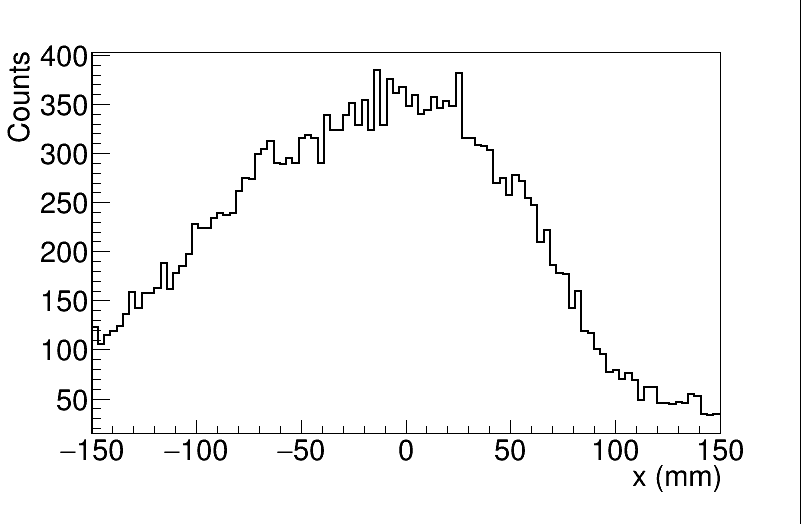}
    \caption{$x$ distribution of counts for measurement (a) corresponding to the $^{137}$Cs source placed at sample location (Fig. \ref{fig:superpose}a)} \label{fig:distria}
\end{figure}

\begin{figure}[!htbp]
\centering
\includegraphics[width=\columnwidth]{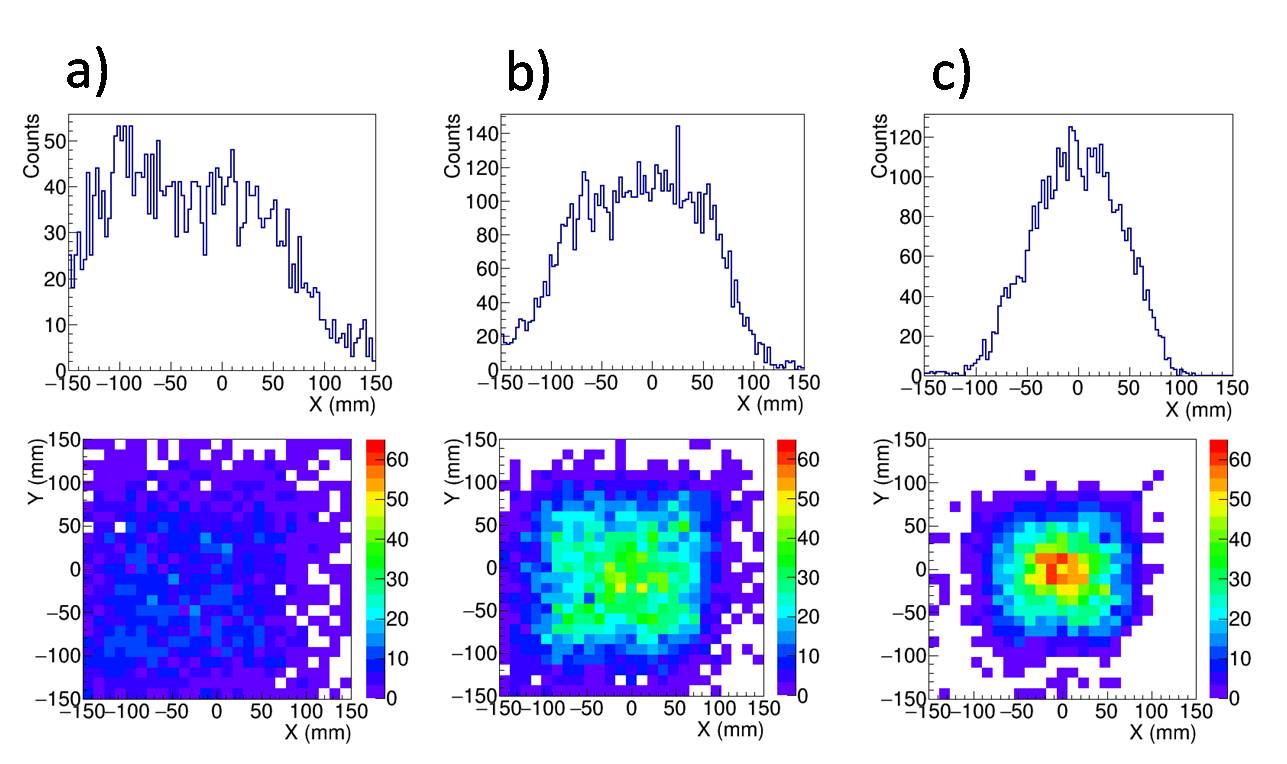}
\caption{Spatial distribution proyected on the x-axis (top) and in the xy-plane (bottom) for backscattering (a), Compton (b) and full-energy events (c) for setup in Fig. \ref{fig:measurements}a} \label{fig:originesa}
\end{figure}

Let us now analyze the results for the measurement (a) corresponding to Fig. \ref{fig:measurements}a in more detail. Its energy spectrum is shown in Fig. \ref{fig:energya}. %We find that the energy resolution at $662$ keV is $9.8\%$ FWHM. 
The $x$ distribution of counts is shown in Fig. \ref{fig:distria}. The aspect that we notice is that it is not symmetric although the $^{137}$Cs source is at the center. In this case we ascribe this effect to $\gamma$-ray backscattering from the Pb-shielding covering the front-surface of the C$_6$D$_6$ detectors. As can be seen in Fig. \ref{fig:superpose}a they are located mainly in the region $x<0$ mm. To investigate this in more detail, we separate the energy spectrum into three different regions: backscattering counts ($E<100$ keV), Compton counts ($100 <E<600$ keV) and photopeak (600~keV $<E<750$~keV) counts. The $x$ distribution and XY map for these three energy-cuts are shown in Fig. \ref{fig:originesa}. As expected, we find now a very symmetric x-distribution for full-energy events. In fact, its width ($\sim100$~mm \textsc{fwhm}) is an estimation of the overall position resolution for points at the center of the focal plane. This also reflects the limited accuracy of the present gamma-imager for spatial cuts. We see that Compton counts have a larger $x$ distribution, which means that they are worse suited than photopeak events to identify the position of the $\gamma$-ray source. Finally, the backscattering events ($E<100$~keV) appear to be spatially more present in the left region ($x<0$ mm), which is consistent with the position of the C$_6$D$_6$ detectors and their lead cover (see Fig.~\ref{fig:measurements}).

%The position resolution at $x=0$ $y=0$ is about XX.

\begin{figure}[!htbp]
\centering
\includegraphics[width=\columnwidth]{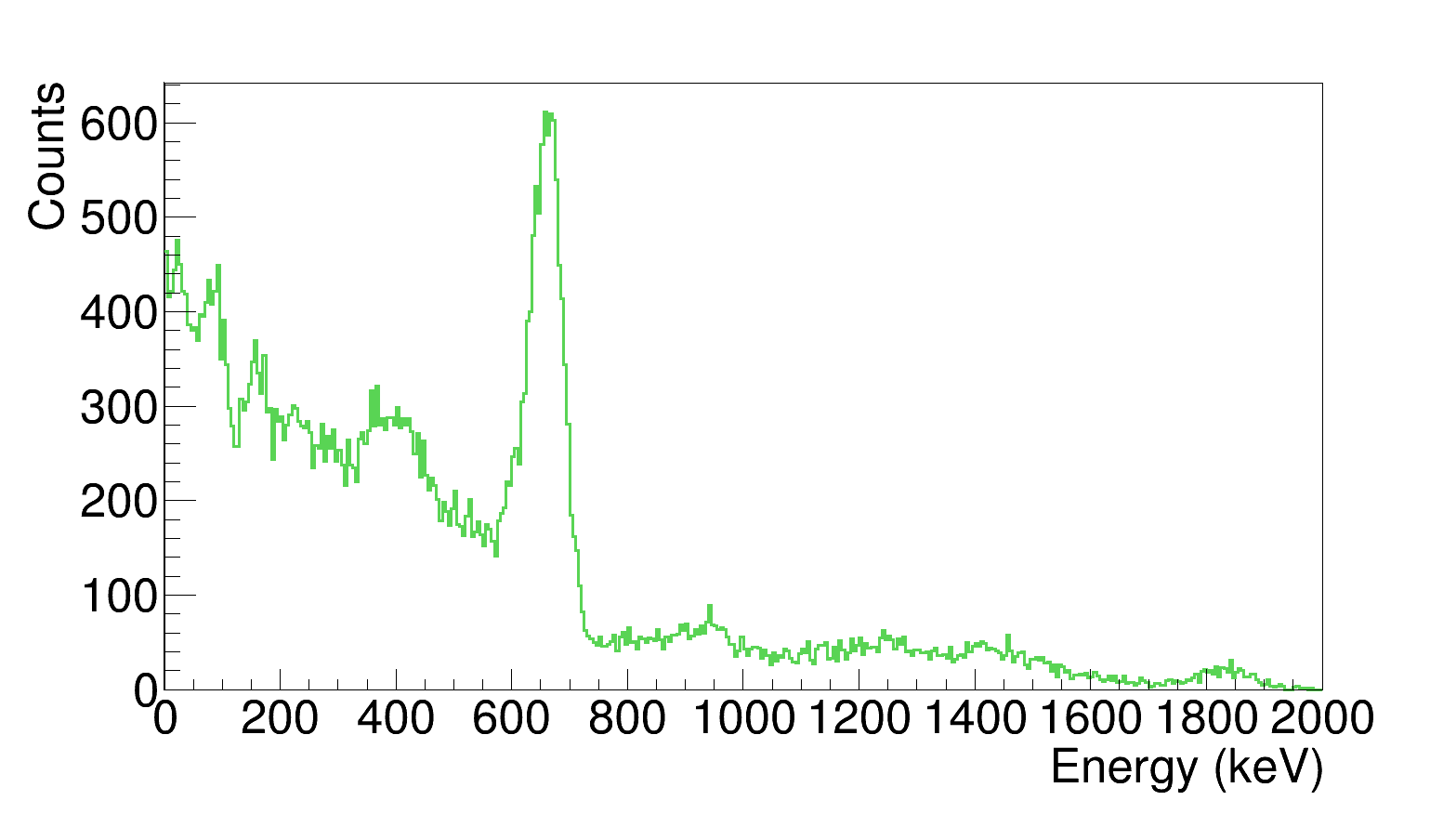}
\caption{Energy spectrum when $^{88}$Y is placed 7 cm upstream and $^{137}$Cs is placed 7 cm downstream (Fig. \ref{fig:measurements}b)} \label{fig:energyb}
\end{figure}

\begin{figure}[!htbp]
\centering
\includegraphics[width=80mm]{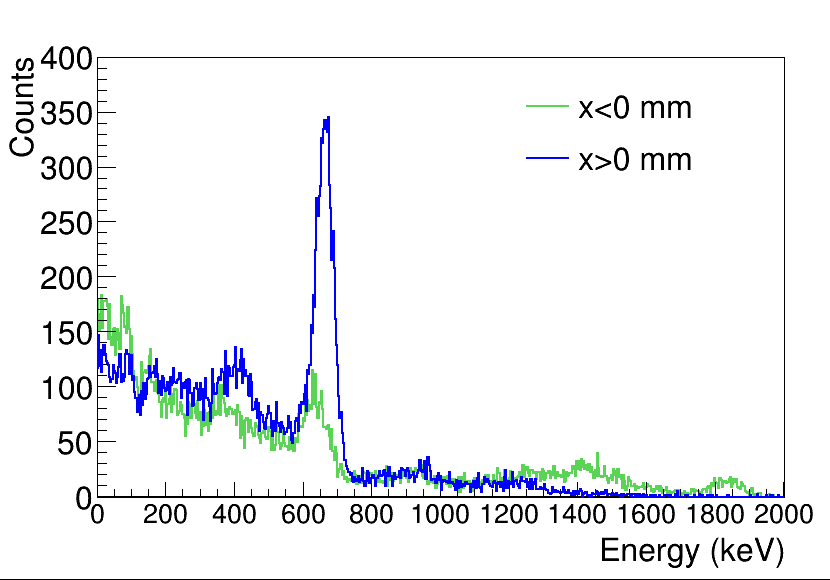}
\caption{Spatial separation of the energy spectrum of $^{137}$Cs and $^{88}$Y.} \label{fig:separation}
\end{figure}

\begin{figure}[!htbp]
\centering
\includegraphics[width=\columnwidth]{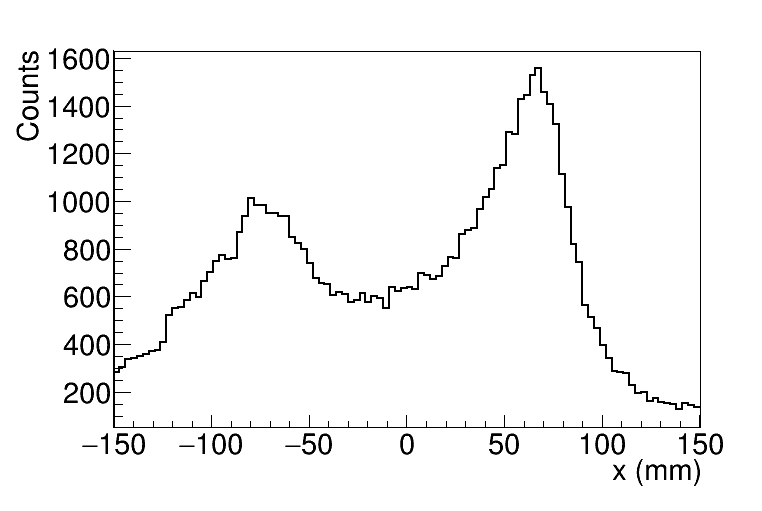}
\caption{$x$ distribution of counts  when $^{88}$Y is placed 7 cm upstream and $^{137}$Cs is placed 7 cm downstream (Fig. \ref{fig:measurements}b)} \label{fig:distrib}
\end{figure}

We turn to measurement (b) corresponding to the setup of Fig.\ref{fig:measurements}b and the gamma-image of Fig.\ref{fig:superpose}b. The energy spectrum in Fig.\ref{fig:energyb} shows the contribution of $\gamma$-rays from both $^{88}$Y- and $^{137}$Cs-sources. It is possible to isolate these spectra to a certain extent by using a spatial criterion. This is demonstrated in Fig. \ref{fig:separation}, where two spectra are built, one formed by events with $x$ positive (region of $^{137}$Cs) and another with $x$ negative (region of $^{88}$Y). In particular we see that the highest photopeak at $1836$ keV only appears in the latter spectrum, confirming that $^{88}$Y  is in the left region. The $x$ distribution is shown in Fig. \ref{fig:distrib}. A similar decomposition to the one in Fig. \ref{fig:originesa} showed us that the main contribution to the region between the two peaks ($-50 < x < 50 $ mm) in Fig. \ref{fig:distrib} is due to Compton counts. 

\begin{figure}[!htbp]
\centering
\includegraphics[width=75mm]{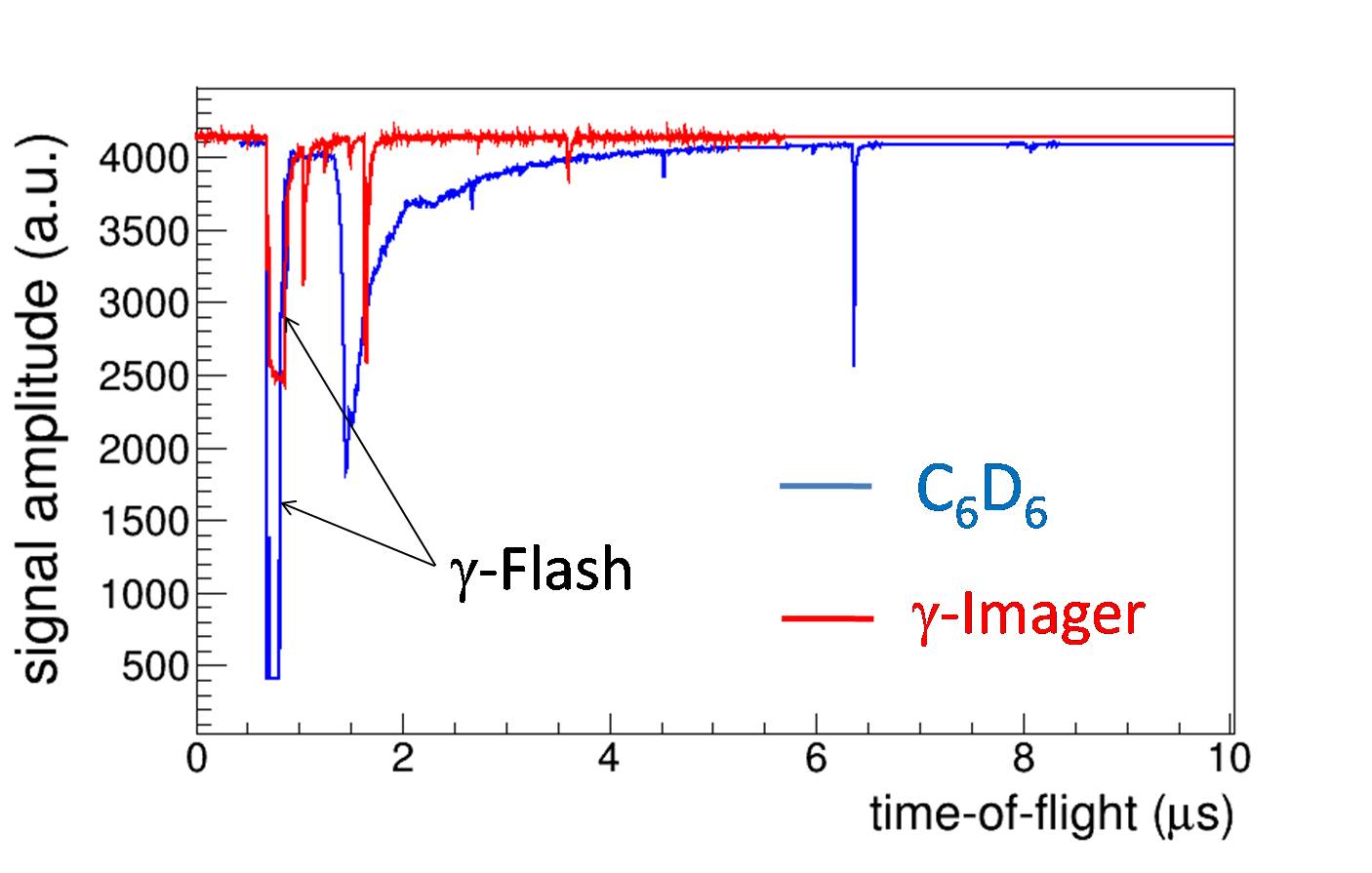}
\caption{Electrical response of the C$_6$D$_6$ and the $\gamma$-ray imager recorded with the digital acquisition system of n\_TOF showing the gamma flash and the following time interval.} \label{fig:flash}
\end{figure}

\begin{figure}[!htbp]
\centering
\includegraphics[width=\columnwidth]{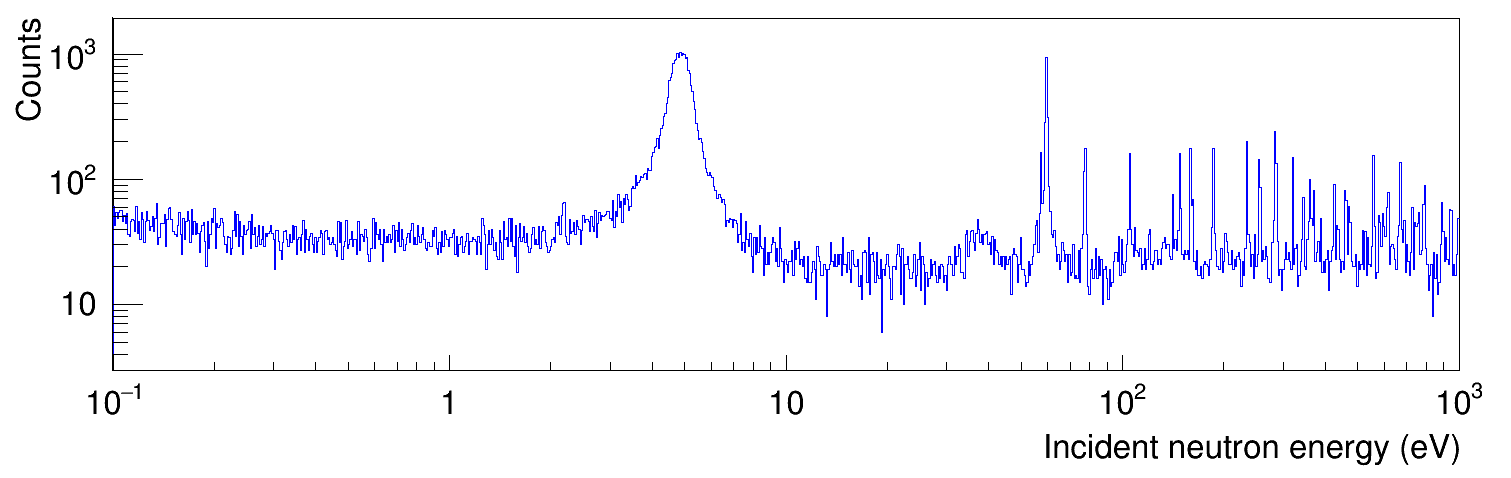}
\caption{Histogram of detected $\gamma$-rays as a function of the incident neutron energy for setup of Fig. \ref{fig:measurements}c.} \label{fig:tofmain}
\end{figure}

We now analyze the neutron capture measurement with the gold sample as shown in the setup of Fig.\ref{fig:measurements}c and the neutron capture measurement of $^{197}$Au ($n,\gamma$) with enhanced background corresponding to the configuration of Fig.\ref{fig:measurements}d. 
The first aspect to emphasize is the raw $\gamma$-imager response, which is compared in Fig. \ref{fig:flash} with the response of a C$_6$D$_6$ detector. The gain recovery of the $\gamma$-imager after the prompt $\gamma$-flash is even faster than that of the C$_6$D$_6$ detectors, an effect which is ascribed to the fast time-response of the CeBr$_3$ crystal and to the smaller efficiency and the large amount of lead-shielding around the position-sensitive detector. This lends confidence on the usability of this type of detector over a relatively broad neutron energy range. In absence of artificial background, the time-of-flight spectrum obtained is displayed in Fig. \ref{fig:tofmain}. It has been expressed in terms of the energy of the incident neutron, by using the conversion\cite{conversion}:
$$ E_n = \left( \frac{72.2977 \cdot L }{t - t_{{flash}}} \right)^2 $$  
where the neutron energy $E_n$ is expressed in units of keV, the time-of-flight $t-t_{flash}$ in $\mu$s, and $L = 183.85$~m is the effective neutron path at n\_TOF. The capture yield is clearly dominated by capture events in $^{197}$Au+n, but a few additional resonances related to the intrinsic neutron sensitivity of the detector are still visible. The main contamination was due to $^{79}$Br and one can clearly identify the $s$-wave resonance at $35.8$ eV\cite{losalamos}.

%Integral evaluation : 2.7E6 à 4E6 (éviter la résonance Br) et 2E7 à 7E7. 

%, and that a high density of resonances associated to this isotope might affect the ones of $^{197}$Au in the region from $E = [50, 1000]$ eV.

%http://journals.aps.org/prc/abstract/10.1103/PhysRevC.81.044616

As expected, a similar TOF-spectrum obtained with any of the four C$_6$D$_6$ detectors shows a better performance in terms of S/B-ratio than the present gamma-camera. This is obvious from the large amount of dead material in the gamma-camera close to the sensitive detection volume, which induces a quite large background. However, it is the aim of this work to explore mainly the capability of implementing $\gamma$-ray imaging as a tool to improve the S/B-ratio of the apparatus itself, which is of relevance for the design of less-massive $\gamma$-ray imaging systems based on electronic collimation~\cite{iTED}.

\begin{figure}[!htbp]
\centering
\includegraphics[width=\columnwidth]{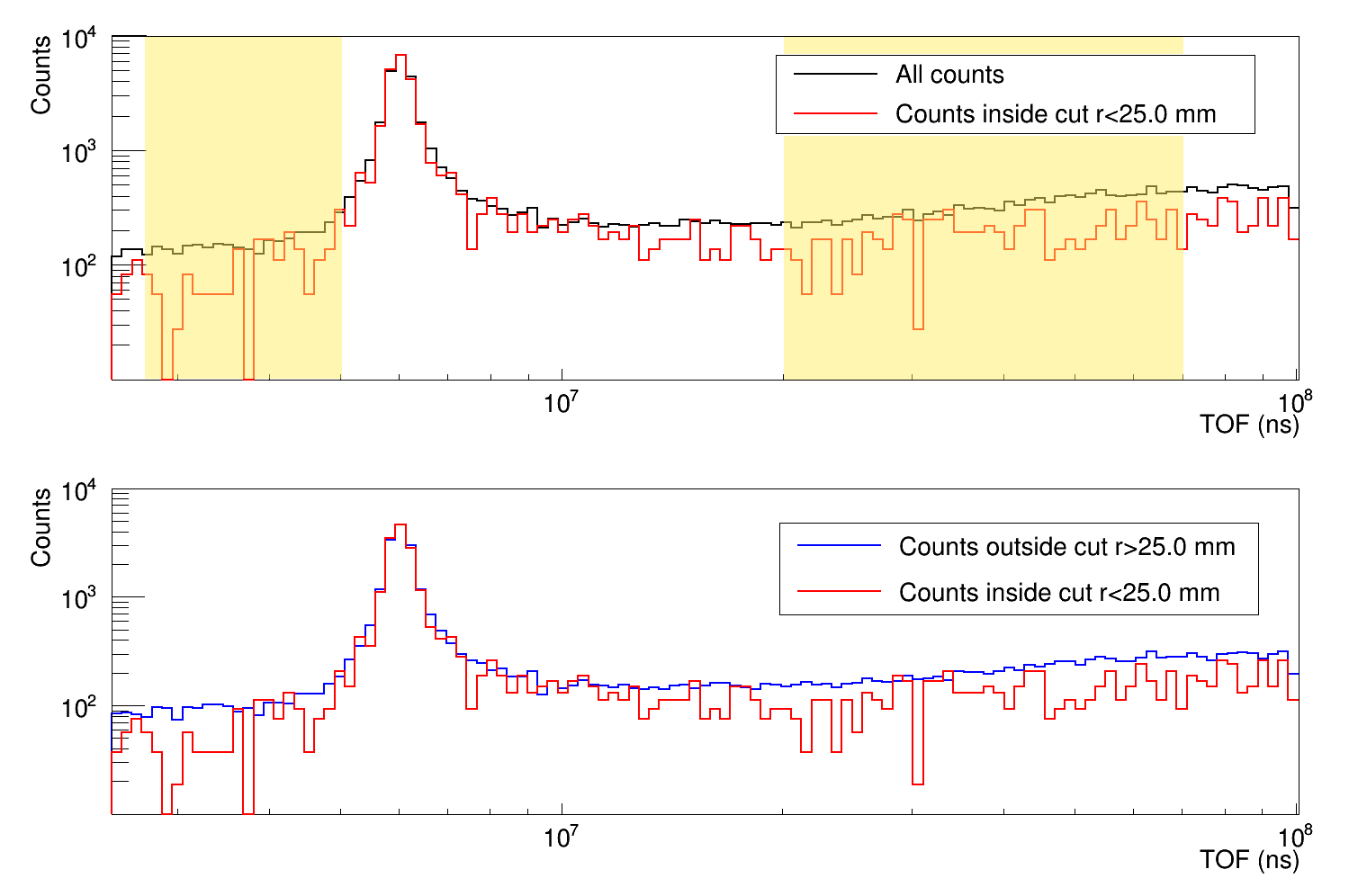}
\caption{Comparison of TOF spectra using a cut of $r=25$~mm obtained with the setup of Fig. \ref{fig:measurements}c. The shaded regions are used for background determination.} \label{fig:twentyfive}
\end{figure}

Thus, we now investigate whether the S/B ratio, defined as the height of a resonance divided by the level of background, can be enhanced by discriminating $\gamma$-rays from another direction than the sample position. To this aim, circular spatial cuts of radius $r$ and centered at $(x,y)=(0,0)$, the sample position, are performed, and the TOF spectra obtained with and without cut are compared. We also compare the spectrum with cut with its complementary, in other terms with events outside the cut. All spectra are normalized to the peak of the 4.9~eV resonance. These spectra for the particular case $r=25$ mm in the region of the first gold resonance at $E_n = 4.9$~eV are shown in Fig. \ref{fig:twentyfive}. We indeed observe a decrease of the background on both sides of the resonance. 

\begin{figure}[!htbp]
\centering
\includegraphics[width=\columnwidth]{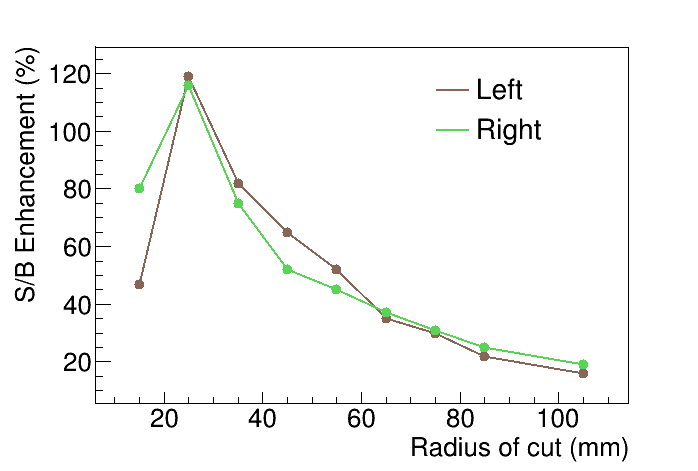}
\caption{Enhancement of the S/B ratio versus circular cuts of radius $r$, for the gold resonance at $E=4.9$ eV.} \label{fig:enhancement}
\end{figure}

To quantify the background level with respect to the peak of the resonance we use the average in the shaded neutron-energy regions indicated in the top panel of Fig. \ref{fig:twentyfive}. The enhancement of the S/B ratio obtained for the different radial cuts plotted in Fig. \ref{fig:enhancement}. The largest improvement obtained (120\%) corresponds to the cut of $r=25$ mm shown in Fig.\ref{fig:twentyfive}. The higher the value of $r$, and hence the size of the spatial cut, the smaller becomes the improvement of the S/B ratio. The enhancements for smaller radial cuts must be considered with caution, because the reduced counting statistics. A similar analysis is carried out for higher energy resonances of $^{197}$Au, at $E=78.2$ and $E=107.0$ eV. For these resonances, the statistics was very low, but we estimate S/B gains respectively up to $30 \%$ and $40 \%$. In both cases the best enhancement is obtained for a radius $r=35$ mm. The lower enhancement in the S/B ratio towards higher neutron energy might be ascribed to the fact that the background starts to be dominated by the neutron sensitivity of the full detection system, mainly from the CeBr$_3$ and the massive collimator.

\begin{figure}[!htbp]
\centering
\includegraphics[width=1.1\columnwidth]{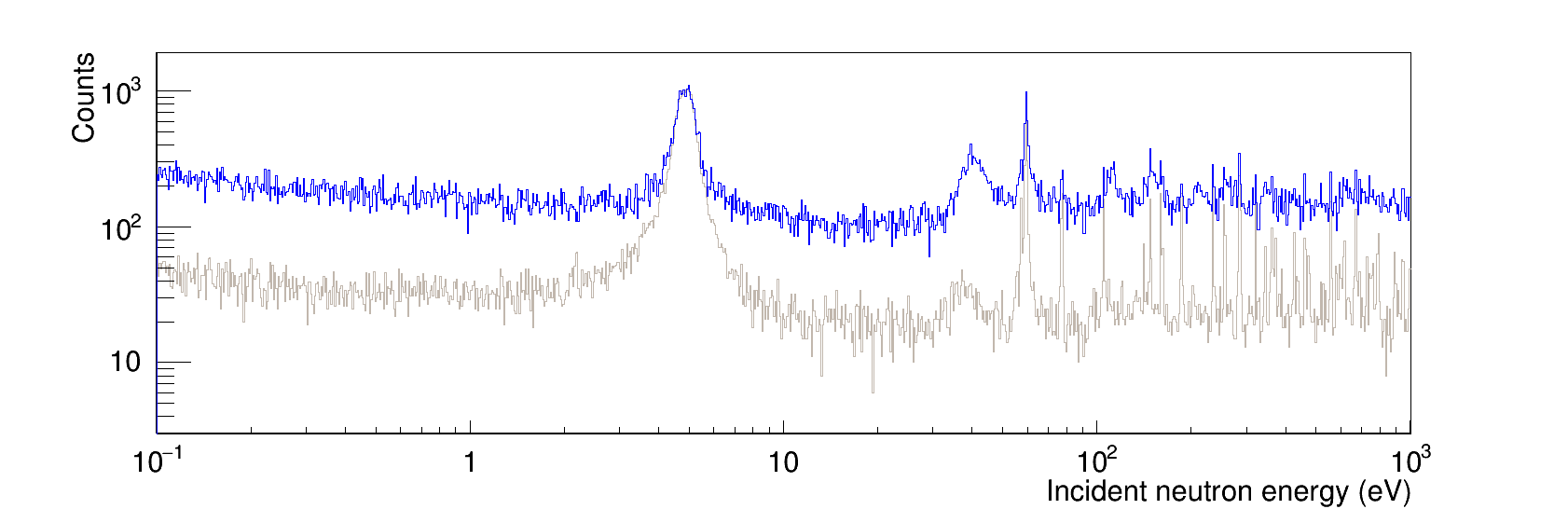}
\caption{Histogram of detected $\gamma$-rays as a function of the incident neutron energy for $^{197}$Au ($n,\gamma$) in presence of artificial background (blue) and without (gray), normalized at $E=4.9$ eV.} \label{fig:tofd}
\end{figure}

We now consider the enhanced background in the TOF spectra for $^{197}$Au($n,\gamma$) for the setup of Fig. \ref{fig:measurements}d with an additional carbon scattering sample in the beam. The $\gamma$-spectrum is shown in Fig. \ref{fig:tofd}, together with the spectrum without artificial background, normalized at $E=4.9$ eV. In this case, the background level increased considerably by almost a factor of five. The dominant  contaminating resonance from $^{71}$Br at $35.8$ keV has increased even by a factor seven. 
%Additional contaminations of $^{79}$Br at energies of $157$~eV and $394$~eV become also more apparent than in measurement (c). 
This strong effect is due to the rather high neutron sensitivity of our gamma camera.  

\begin{figure}[!htbp]
\centering
\includegraphics[width=\columnwidth]{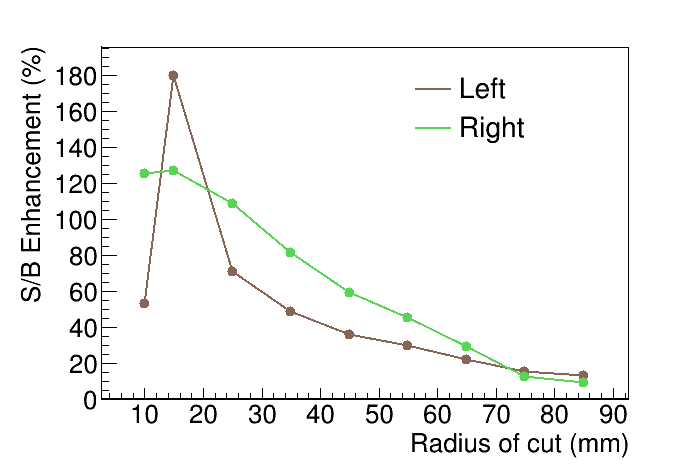}
\caption{Enhancement of S/B ratio for different circular cuts of radius $r$, with respect to the situation where all events are taken into consideration, for the resonance $E=4.9$ eV.} \label{fig:enhancementbis}
\end{figure}

\begin{figure}[!htbp]
\centering
\includegraphics[width=\columnwidth]{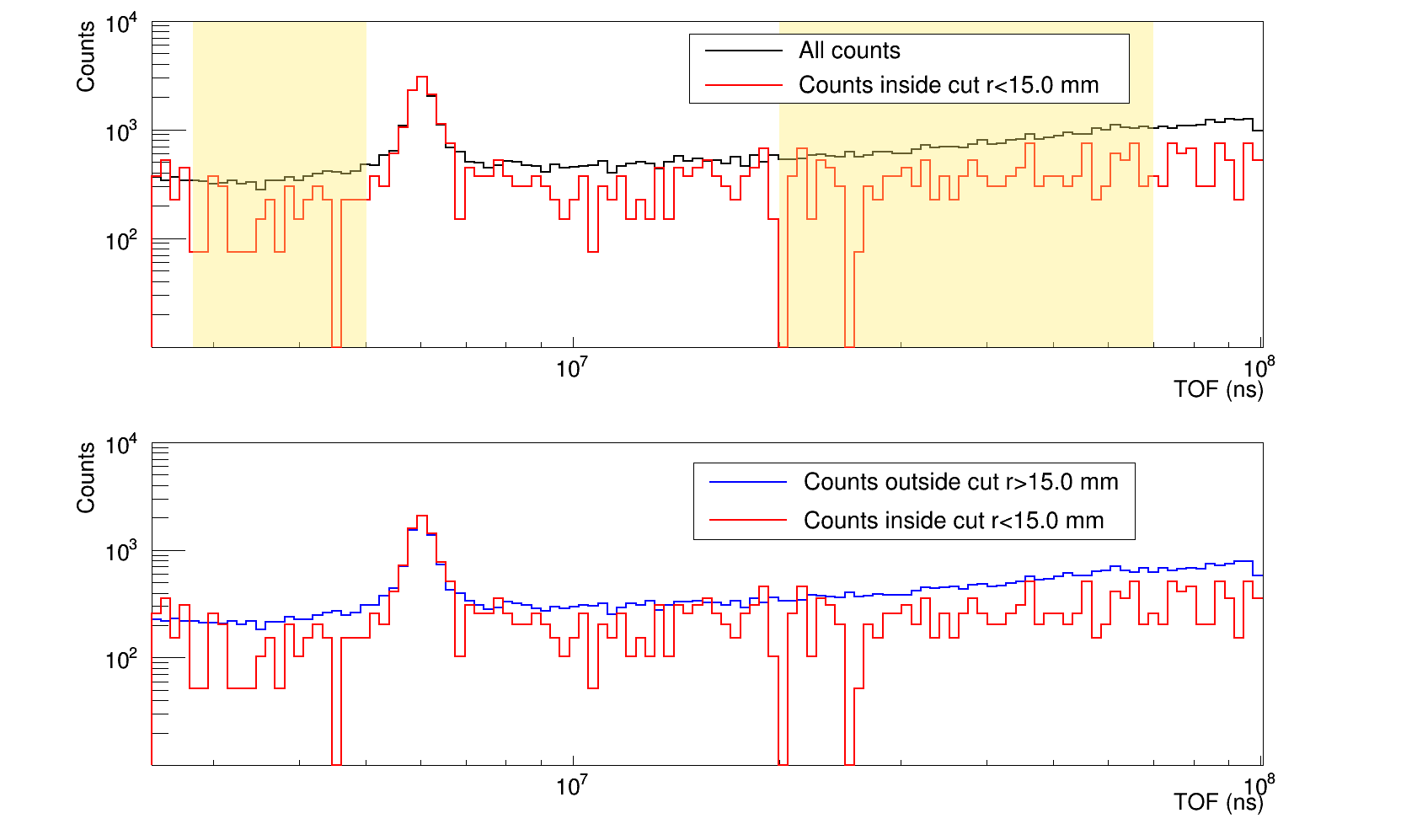}
\caption{Comparison of TOF spectra obtained with the setup of Fig. \ref{fig:measurements}d and a cut $r=15$~mm. The shaded regions were used for background determination.} \label{fig:fifteen}
\end{figure}

Once again, circular cuts are implemented and the gain in S/B ratio is illustrated in Fig.~\ref{fig:enhancementbis}. We obtain enhancements between 120\% and $180\%$, which show that even under severe background conditions our device is working properly and that the detection sensitivity of the sample-related capture channel can be enhanced by means of $\gamma$-ray imaging. We observe now a larger S/B improvement on both sides of the 4.9~eV resonance, when compared to the results obtained with the setup of Fig.~\ref{fig:measurements}c. We ascribe this difference to the stronger TOF dependency of the background by both radioactive sources and the carbon sample, which contribute to the background mainly on the right-hand side (large TOF values) of the resonance. The best cut is in this situation $r=15$~mm. A similar comparison of the TOF spectra in absence of artificial background is shown in Fig. \ref{fig:fifteen} for $r=15$ mm. We also studied the resonances at $E=78.2$ and $122$~eV, again with correspondingly lower statistics. The best enhancements of the S/B ratio are $50$ and $70\%$ respectively, in both cases for a radius $r=35$ m. The limited statistics prevent us from extending our analysis to higher neutron energies, were the capture levels of gold are obscured by the low statistics and by $^{79}$Br contaminations.

\section{Conclusions and outlook}~\label{sec:outlook}
The aim of this work was to study the applicability of $\gamma$-ray imaging techniques as a possible tool for background rejection in ($n,\gamma$) measurements. To this aim, we developed and implemented a pinhole gamma camera. Although not optimal for ($n,\gamma$) measurements given the large amount of dead material, this apparatus allowed us to determine the direction of incident gamma radiation and to study the possibility for discriminating capture $\gamma$-rays from the sample from contaminant $\gamma$-rays from the surroundings. This option was investigtaed in specific measurements at CERN n\_TOF using calibrated radioactive sources and in a series of $^{197}$Au ($n,\gamma$) measurements. The detector response and the measured $\gamma$-ray spectra as a function of neutron energy demonstrate the proper performance of the full system in a real neutron capture TOF experiment. Improvements in signal-to-background ratio of up to a factor of two were achieved by implementing spatial cuts in the data analysis. This result is quite promising considering the limitations of the instrumentation employed, mainly the high intrinsic neutron sensitivity and the limitations concerning the field of view and in the position resolution. The present measurements also serve to confirm the proper performance of the detector components, such as the pixelated photomultiplier or the AMIC2GR readout system. 

The next step will be to implement a $\gamma$-ray imager based on the Compton imaging technique, which can be implemented with much less dead material and shows a better performance in terms of angular resolution and efficiency. Also it is preferred to employ inorganic crystals with lower intrinsic neutron sensitivity, such as LaCl$_3$, similar to the i-TED modules described in \cite{iTED}. Such a system should allow us to enhance the field of view and the accuracy of the spatial cuts, thus enabling a higher detection efficiency and more stringent angular cuts for a superior background rejection. A comparison with state-of-the-art C$_6$D$_6$ detectors will be also carried out in order to benchmark the overall performance of the proposed methodology and instrumentation.

\section*{Acknowledgments}
We thank I3M at Universidad Polit\'ecnica de Valencia (Spain) for lending us the ASIC AMIC2GR device used in this work. 
This work was partially supported by Spanish \textit{Ministerio de Econom\'{\i}a y Competitivdad} under Grants No. FPA2011-24553 and FPA2013-45083-P. CDP acknowledges finantial support from a grant of the \textit{Generalitat Valenciana GV2013 Grupos de Investigaci\'on Emergentes} and a grant from \textit{Universidad de Valencia UV-INV-PRECOMP12-80717}. The authors thank B.~Morse for his useful commments and corrections.
%\section*{References}
%\bibliographystyle{unsrt} % Le style est mis entre accolades.
\bibliography{referencias}

\end{document}